\documentclass[aps,preprint,amsmath,amssymb,citeautoscript,showpacs]{revtex4}

\newcommand{\psib}{{\overline{\psi}}}

\usepackage{graphicx,times}

\begin{document}

\title{The fermion bag approach to lattice field theories}
\author{Shailesh Chandrasekharan}
\affiliation{Department of Physics, Box 90305, Duke University,
Durham, North Carolina 27708, USA\thanks{permanent address} \\ and \\
Department of Theoretical Physics, Tata Institute of Fundamental Research, 
Homi Bhaba Road, Mumbai 400005, India}

\begin{abstract}
We propose a new approach to the fermion sign problem in systems where there is a coupling $U$ such that when it is infinite the fermions are paired into bosons and there is no fermion permutation sign to worry about. We argue that as $U$ becomes finite fermions are liberated but are naturally confined to regions which we refer to as {\em fermion bags}. The fermion sign problem is then confined to these bags and may be solved using the determinantal trick. In the parameter regime where the fermion bags are small and their typical size does not grow with the system size, construction of Monte Carlo methods that are far more efficient than conventional algorithms should be possible. In the region where the fermion bags grow with system size, the fermion bag approach continues to provide an alternative approach to the problem but may lose its main advantage in terms of efficiency. The fermion bag approach also provides new insights and solutions to sign problems. A natural solution to the ``silver blaze problem'' also emerges. Using the three dimensional massless lattice Thirring model as an example we introduce the fermion bag approach and demonstrate some of these features. We compute the critical exponents at the quantum phase transition and find $\nu=0.87(2)$ and $\eta=0.62(2)$.
\end{abstract}

\pacs{71.10.Fd,02.70.Ss,11.30.Rd,05.30.Rt}

\maketitle

\section{Introduction}

Theories containing fermions at a microscopic level which interact strongly with each other are of interest in both condensed matter and particle physics. In condensed matter physics, such theories are used to describe quantum critical behavior in strongly correlated electronic materials \cite{Sachdev:2008,Gegenwart:2008}.  Strongly interacting gapless Dirac fermions arise naturally in the physics of graphene \cite{neto-2009-81}. In particle physics, the $u$ and $d$ quarks which are almost massless, interact with each other strongly to produce the complex dynamics of nuclear physics \cite{Durr:2008zz,Savage:2005ma}. Even the Higgs particle of the standard model, which remains to be discovered, could be a strongly coupled bound state of fermionic particles that may exist beyond the standard model \cite{RevModPhys.55.449}.

Although the microscopic theory for these wide range of phenomena are quite different, the computational difficulties that one encounters when dealing with strongly interacting fermions are remarkably similar. Firstly, perturbative methods are not applicable since there are no small parameters in the problem. Other methods, which go beyond perturbation theory, such as mean field theory, often involve uncontrolled approximations. The best alternative approach is the Monte Carlo method. However, the state of the art of Monte Carlo methods for fermionic systems is still primitive. The major stumbling block is the infamous fermion sign problem, which arises due to the quantum nature of fermions and needs to be solved before importance sampling techniques can be employed. In this context it is useful to distinguish fermion sign problems with other sign problems that arise in lattice field theories. For example in some formulations, even bosonic lattice field theories contain sign problems in the presence of a chemical potential that favor particles over anti-particles \cite{aarts:131601}. However, these sign problems are solvable completely in a different formulation \cite{Endres:2006xu,Chandrasekharan:2008gp}. There are indeed sign problems in bosonic lattice field theories that remain unsolvable. These arise when bosons interact with gauge fields in the presence of a chemical potential or contain frustrations \cite{Nyfeler:2008ck}. In this work we focus on the fermion sign problem although some of the ideas may be applicable more generally.

Solutions to sign problems always involve re-summation over a class of configurations. This re-summation is cumbersome and makes the Monte Carlo updates slow. Two methods have been discovered so far to solve the fermion sign problem completely. One is the auxiliary field method \cite{PhysRevB.34.7911}, and the other is the meron cluster method \cite{PhysRevLett.83.3116}. The auxiliary field method is based on converting an interacting fermion problem into a free fermion problem in the background of an auxiliary field. The sum of all free fermion configurations is equal to the determinant of a fermion matrix. If this determinant can be shown to be always positive the sign problem is solved. We refer to Monte Carlo algorithms based on of this approach as conventional algorithms. Even when the sign problem is solved, these conventional algorithms can be inefficient since the problem becomes completely non-local in the system size. One well known problem is that often the fermion matrix develops a large number of small eigenvalues. In these cases the algorithms slow down substantially with system size. In practical calculations, the small eigenvalues of the fermion matrix are controlled by the addition of new couplings to the theory which are then extrapolated to zero to extract physical answers. This introduces systematic errors which cannot easily be controlled. Finally, and most importantly, when the determinant is not positive, little insight can be gained about the fermion sign problem itself. In contrast to the auxiliary field method, the meron cluster method is based on cleverly rewriting the partition function as a sum over configurations that naturally divide the physical system into clusters or regions so that the sign problem is solved by re-summing configurations within each region. Due to the cleverness involved, the method is not widely applicable. On the other hand whenever it works, large system sizes can be studied more easily since the problem breaks up the system into smaller regions and one does not have to consider the entire system size to solve the fermion sign problem. In particular lattices of the order of $128 \times 128$ have been solved using this method \cite{PhysRevB.66.045113}. Additional couplings to control the efficiency of the algorithm become unnecessary.

In this work we propose a more general approach to the fermion sign problem based on the underlying physics. In a sense we extend the meron cluster idea by combining it with the determinantal trick to solve the fermion sign problem in a wider class of theories. The essential idea is that many fermionic theories contain a coupling, which we will call $U$, such that when $U=\infty$ the fermions become paired into bosons and the partition function is naturally written with positive definite Boltzmann weights. In other words there is no fermion sign to worry about. When the coupling is large but not infinite, fermions become unpaired but remain confined to small regions which we refer to as {\em fermion bags}. The fermion sign problem is then confined to these bags and can sometimes be solved using the usual determinantal trick. When the bags remain small, the computational effort to solve the sign problem does not grow with the system size just like the meron cluster approach. Thus, Monte Carlo methods for these problems can be far more efficient than algorithms which do not take this physics into account. As the coupling reduces further the fermion bags merge and begin to grow with the volume. In this region the fermion bag approach loses its main advantage and suffers form similar slowing down as the auxiliary field methods. However, it is useful to remember that at small couplings perturbation theory is usually a good approach and the recently proposed diagrammatic Monte Carlo method may be a better approach for small and moderate values of the couplings \cite{prokof'ev:250201}.

The main message behind the fermion bag idea is the following: {\em When fermions are delocalized over the whole system, the increased computational cost associated to dealing with fermionic degrees of freedom is natural. But it is definitely unnatural in the regime where fermions are confined to small regions.} The auxiliary field method to the fermion sign problem does not make use of this underlying physical picture. The similarity of the fermion bag approach to the meron cluster approach is striking: The bags, like the clusters, do not occupy the whole volume and makes the computational effort somewhat reduced. In addition, as we will discuss in this work, new insights and solutions to the fermion sign problems emerge. The fermion bag idea was first discussed in \cite{Chandrasekharan:2008gp}.

Our article is organized as follows. In section 2 we illustrate the ideas outlined above concretely using a simple but relatively less studied example of the massless lattice Thirring model constructed with a single flavor of staggered fermions. In particular we contrast the fermion bag approach with the auxiliary field method. In section 3, we introduce a fermion chemical potential and discuss how the silver-blaze problem \cite{PhysRevLett.91.222001}, present in the auxiliary field method, is naturally solved in the fermion bag approach. In section 4, we given an example of a sign problem which seems unsolvable in the auxiliary field formulation but is solvable in the fermion bag approach. In Section 5 we discuss update algorithms for the massless Thirring model in the bag formulation. In Section 6, we discuss the fermion bag distribution in the massless Thirring model. In particular we show that the typical fermion bag size does not grow with system size for $U \gtrsim 1.2$. In section 7, we discuss some results obtained using the bag approach in the massless Thirring model and in section 8, we discuss the quantum critical behavior. Section 9 contains our conclusions where we argue that the fermion bag approach is rather general and must be applicable to many lattice field theories. In particular we show how similar ideas may be adapted to the physics of the BCS-BEC crossover.

\section{The Fermion Bag Approach}

Although the fermion bag approach is applicable to a wide class of problems in any dimension, it is useful to understand the details in the context of a simple model. Here we introduce the fermion bag approach using the example of the massless lattice Thirring model with one flavor of staggered fermions on a three dimensional cubic lattice. The action is given by 
\begin{equation}
S = - \sum_{x,y} {\overline\psi}_x \ D_{x,y} \ \psi_y
\ - U \ \sum_{x,\alpha} 
{\overline\psi}_{x+\hat\alpha}\psi_{x+\hat\alpha} {\overline\psi}_x\psi_x
\label{act}
\end{equation}
where the matrix $D$ is the free staggered Dirac operator given by \cite{Sharatchandra:1981si}
\begin{equation}
D_{x,y} = \frac{\eta_{x,\alpha}}{2} 
[\delta_{x+\hat{\alpha},y}-\delta_{x,y+\hat{\alpha}}].
\label{staggered}
\end{equation}
In our notation $x \equiv (x_1,x_2,x_3)$ denotes a lattice site on a $3$ dimensional cubic lattice of size $L^3$, $\overline\psi_x$ and $\psi_x$, are Grassmann valued fields and $\alpha = 1,2,3$ runs over the three positive directions. The staggered fermion phase factors $\eta_{x,\alpha} = \exp(i\pi \zeta_\alpha \cdot x)$ are defined through the 3-vectors $\zeta_1 = (0,0,0), \zeta_2=(1,0,0)$ and $\zeta_3=(1,1,0)$. We also define the phase $\varepsilon_x = (-1)^{x_1+x_2+x_3}$ for later convenience.

The main feature of the model is that it contains massless fermions interacting with each other with a $U_f(1) \times U_\chi(1)$ chirally invariant interaction. Indeed it is easy to check that the action is invariant under the usual fermion number $U_f(1)$ transformations: $\psi_x \rightarrow \exp(i\theta) \psi_x$ and $\psib_x \rightarrow \exp(-i\theta) \psib_x$, and the chiral $U_\chi(1)$ transformations: $\psi_x \rightarrow \exp(i\varepsilon_x \theta) \psi_x$ and $\psib_x \rightarrow \exp(i\varepsilon_x \theta) \psib_x$. When $U=0$ the model describes free massless Dirac fermions. At infinite $U$, all fermions are confined and the model reduces to a hardcore dimer model made up of paired fermions and the low energy physics is in the same universality class as the $XY$ model in its broken phase \cite{Chandrasekharan:2003qv}. Hence at some critical value $U_c$ the model undergoes a quantum phase transition. This model and its variants have been studied earlier with the auxiliary field method \cite{PhysRevLett.59.14,Shigemitsu:1991xu,AliKhan:1993dx,DelDebbio:1995zc,Debbio:1997dv,Barbour:1998yc,Debbio:1999xg,Hands:1999id,Christofi:2007ye}. However, none of the earlier calculations were performed in the massless limit due to algorithmic difficulties. Here we use the fermion bag approach to tackle the massless limit for the first time.

The partition function of the model is given by
\begin{equation}
Z = \int \prod_x [d\psi_x\ d\overline\psi_x] \ \exp(-S)
\label{modpf}
\end{equation}
where the integration is over the Grassmann fields. In the determinantal approach one uses the Hubbard-Stratanovich transformation to convert the four fermion coupling into a fermion bi-linear at the cost of introducing an integral over an auxiliary bosonic field. It is easy to verify that
\begin{equation}
Z = \int \ d\phi \  [d\psib d\psi] \ \exp\Big\{\ 
\sum_{x,y} \psib_x (M[\phi])_{x,y} \psi_y \Big\}
\end{equation}
where $M[\phi]$ is given by
\begin{equation}
M([\phi]) = \eta_\mu(x) \Big[\delta_{x+\mu,y}(\frac{1}{2}+ \sqrt{U}\mathrm{e}^{i\phi_\mu(x)}) - \delta_{x,y+\mu}(\frac{1}{2}+\sqrt{U}\mathrm{e}^{-i\phi_\mu(x)})\Big].
\end{equation}
The auxiliary field $\phi_\alpha(x)$ is integrated over the angles $0 \leq \phi_\mu(x) < 2\pi$. Integrating over the Grassmann variables first we can obtain
\begin{equation}
Z = \int [d\phi] \ \mathrm{Det}(M([\phi])).
\label{convpf}
\end{equation}
The matrix $M$ is anti-Hermitian and so its eigenvalues are purely imaginary. Further, it anti-commutes with the matrix $\Xi_{x,y} = \epsilon_x \delta_{x,y}$ which means that if $\lambda$ is an eigenvalue then so is $-\lambda$. Thus, $\mathrm{Det}(M[\phi]) \geq 0$ for every $[\phi]$ and the sign problem is solved. While different Monte Carlo algorithms exist to solve the remaining problem, the most popular is the Hybrid Monte Carlo (HMC) method  due to its favorable scaling with the volume \cite{PhysRevB.36.8632}.

Let us now briefly discuss the cost of the HMC algorithm. The HMC method is based on generating a new independent configuration $[\phi]$ based on a series of molecular dynamics update. The new configuration is then accepted or rejected based on a Metropolis accept reject step. Let $N_{\mathrm{MD}}$ be the number of molecular dynamics steps necessary to generate a statistically independent configuration. Each step of the molecular dynamics update requires the computation of the force which requires a particular matrix element of $(M[\phi])^{-1}$. Typically this requires $N_{\mathrm{CG}} L^3$ operations where $N_{\mathrm{CG}}$ is the number of conjugate gradient steps in the inversion process. Thus, the cost of generating an independent configuration in an HMC is given by $L^3 N_{\mathrm{CG}} N_{\mathrm{MD}}$. Both $N_{\mathrm{CG}}$ and $N_{\mathrm{MD}}$ are dependent on the physics and the model. In the current context, for large and intermediate values of $U$, the matrix $M[\phi]$ contains a non-zero density of small eigenvalues due to chiral symmetry breaking. Hence one expects $N_{\mathrm{CG}} \sim L^3$. On the other hand $N_{\mathrm{MD}}$ grows with the largest correlation length in the problem and for the moment we will assume this to be $L$ which is the best case scenario since the theory contains massless particles. Thus, the HMC effort scales at least as $L^7$. One can reduce $N_{\mathrm{CG}}$ drastically if we can can control the small eigenvalues of the matrix $M[\phi]$. This is usually accomplished by adding a fermion mass term. This is the reason why all previous calculations of the Thirring model always used a non-zero fermion mass. No calculations of the massless Thirring model at large $U$ have been attempted using the HMC method. At small values of $U$ experience shows that $N_{\mathrm{CG}} \sim L$ since the fermions are almost free. Assuming that again $N_{\mathrm{MD}} \sim L$, the HMC effort now scales as $L^5$.

How is the fermion bag approach different? Instead of introducing an auxiliary field to rewrite the four-fermion term as a fermion bi-linear, we begin with the partition function given by
\begin{equation}
Z = \int [d\psi d\psib] \exp\Bigg(
\Big\{ \sum_{x,y} \psi_x D_{x,y} \psi_y
+ \sum_{x,\alpha}  U \psib_x\psi_x\psib_{x+\alpha}\psi_{x+\alpha}\Big\}\Bigg),
\end{equation}
and expand it in powers of $U$ using
\begin{equation}
\exp(U \psib_x\psi_x\psib_{x+\alpha}\psi_{x+\alpha}) = 
1 + U \psib_x\psi_x\psib_{x+\alpha}\psi_{x+\alpha}.
\end{equation}
The Grassmann integration then gives
\begin{equation}
Z = \sum_{n_{x,\alpha}=0,1} \Big(\prod_{x,\alpha} U^{n_{x,\alpha}}\Big) 
\mathrm{Det}(W[n])
\label{fbpf}
\end{equation}
where $n_{x,\alpha}=0,1$. The $n_{x,\alpha}=1$ bonds are referred to as dimers. Note that in this approach the Grassmann integration leads to a determinant of a different matrix $W[n]$, which is just the free fermion matrix where the sites connected to $n_{x,\alpha}=1$ are dropped. It is easy to verify that $W[n]$ is also anti-Hermitian and anti-commutes with $\Xi$ and so $\mathrm{Det}(W[n]) \geq 0$. Thus the sign problem is again solved.

\begin{figure}[t]
\begin{center}
\includegraphics[width=0.7\textwidth]{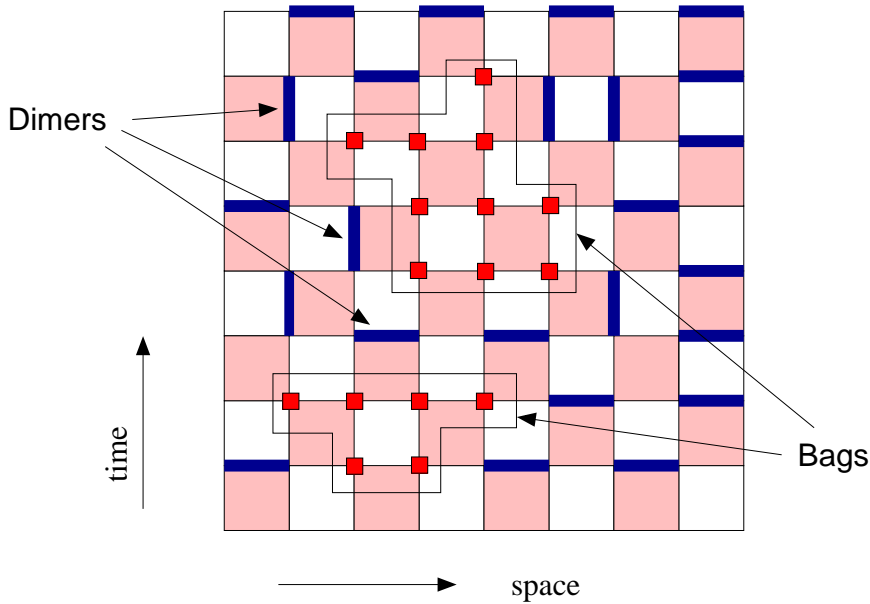}
\end{center}
\caption{\label{fig1} An illustration of a ``fermion-bag'' configuration as discussed in the text.}
\end{figure}

Let us now show that we have captured important physics in this new formulation. Note that the configuration $[n]$ divides the lattice into disconnected regions or ``bags'' ${\cal B}_i,i=1,2,...$. Each bag consists of sites connected with only $n_{x,\mu}=0$ bonds. Inside each bag the fermions hop freely while outside they are confined in the form of dimers. The size and shape of the bags are dynamically determined by the value of $U$. One such configuration is illustrated in Fig.~\ref{fig1}. Note that a single world line configuration of fermions inside the bag can give negative weights due to quantum mechanics. However, we can resum all the possible fermion world lines within the bag exactly. Indeed the quantum interference of all the fermion paths inside the bag ${\cal B}_i$is simply $\mathrm{Det}(W[{\cal B}_i]) \geq 0$ and so
\begin{equation}
\mathrm{Det}(W[n]) = \prod_i \mathrm{Det}(W[{\cal B}_i])
\end{equation}
Thus, we see that fermions have become classical objects when they are considered as non-local objects in the form of bags. For this reason we call our method as the {\em fermion-bag} approach. The size, the shape and other properties of these bags encode the fermion physics.

Let us now discuss the effort required to generate a statistically independent configuration in the fermion bag approach using a specific algorithm discussed later in section \ref{MC}. In our algorithm, a local update requires the computation of a single matrix element of $(W[{\cal B}_x])^{-1}$ where ${\cal B}_x$ is the bag associated to the site $x$. The effort associated with this step is equal to $N_{B} N_{CG}$. The total effort of obtaining a statistically independent configuration is then of the order $N_B N_{CG} L^3$. Here we assume that one sweep through the lattice is sufficient to generate such a configuration. This is almost true due to the availability of a directed path update (see section \ref{MC}).  When $U$ is large, the bags are small comprising of a few neighboring bonds and thus independent of the volume. Although each local update can be difficult the computational cost of a local update ($N_B N_{CG}$) does not grow with the volume. This makes the fermion bag approach far more efficient for large system sizes compared to the determinantal approach. The former scales as $N_B N_{CG} L^3$ while the latter scales as $L^7$ as discussed earlier.

When $U$ is small the bags can percolate and become as big as the system size. Here we expect $N_B \sim L^3$. On the other hand since the fermions are almost free, the matrix inversion using the conjugate gradient algorithm becomes easy. we find that $N_{CG} \sim L$ and hence the overall effort now grows as $L^7$. On the other hand the auxiliary field method based on the HMC algorithm scales as $L^5$ and so is clearly superior. It may be interesting to explore a HMC type algorithm in the fermion bag approach if possible so as to combine the good features of both. But this is not the focus of the current work. Further, as mentioned earlier, at small $U$ the diagrammatic Monte Carlo algorithm may be the best option \cite{prokof'ev:250201}.

 At intermediate values of $U$, especially close to the phase transition, the HMC most likely continues to scale as $L^7$ or worse due to critical slowing down. On the other hand the scaling of the bag algorithm is more tricky and needs to be studied carefully. We find three reasons to remain optimistic: (1) The bags of all sizes exist in the simulation so some updates are much faster, (2) The matrix $W{\cal B}_x$ is the free matrix except for mesoscopic fluctuations coming from the boundaries of the bag. Hence $N_{CG}$ may scale favorably in the bag approach, (3) The existence of the directed path algorithm to update variables outside the bag may eliminate a lot of the critical slowing down. More research is necessary to compare the two algorithms in the intermediate $U$ region. 

\section{Solution of the Silver Blaze Problem}

The fermion bag approach also offers new insights into sign problems. Here we discuss a simple resolution of the so called {\em silver blaze problem}, a general paradox related to sign problems that arises in the auxiliary field method in the presence of a fermion chemical potential \cite{PhysRevLett.91.222001}. Before we discuss how the fermion bag approach solves this problem, let us first review its origin in the current context of the massless lattice Thirring model. 

In the presence of a chemical potential $\mu$, the Dirac operator given in Eq.~(\ref{staggered}) changes to
\begin{equation}
D(\mu)_{x,y} = \frac{\eta_{x,\alpha}}{2} 
[\delta_{x+\hat{\alpha},y}\mathrm{e}^{\mu\delta_{\alpha,t}}
-\delta_{x,y+\hat{\alpha}}\mathrm{e}^{-\mu\delta_{\alpha,t}}].
\label{Dmu}
\end{equation}
In the auxiliary field method the four-fermion term is again converted to a fermion bi-linear using the Hubbard Stratanovich transformation and the partition function is again given by Eq.~(\ref{convpf}), except that the matrix $M[\phi]$ is now given by
\begin{equation}
M([\phi]) = \eta_\alpha(x) \Big[\delta_{x+\mu,y}(\frac{1}{2}+ \sqrt{U}\mathrm{e}^{i\phi_\alpha(x)+\mu \delta_{\alpha,t}}) - 
\delta_{x-\mu,y}(\frac{1}{2}+\sqrt{U}\mathrm{e}^{-i\phi_\mu(x)-\mu\delta_{\alpha,t}})\Big].
\label{muM}
\end{equation}
Unfortunately, the properties that we used to argue that $\mathrm{Det}(M[\phi]) \geq 0$ are no longer valid when $\mu \neq 0$. Indeed the determinant can be negative as soon as $\mu \neq 0$ for all values of $U$. This is the well known sign problem in the presence of a chemical potential.

Consider large values of $U$ where the fermions become massive due to chiral symmetry breaking. In this phase the chemical potential should have no effect on the ground state of the system until a critical chemical potential is reached. This means, for low temperatures a small chemical potential must have little effect on the physics. However, the sign problem does not respect this mild behavior with respect to the chemical potential. The sign problem becomes severe as soon as the chemical potential is non-zero at small temperatures. This paradox has been called the {\em silver-blaze problem} \cite{PhysRevLett.91.222001}. The auxiliary field method offers almost no explanation for this paradox, except the fact that the cancellations due to the sign problem are crucial to get the right physics.

\begin{figure}[t]
\begin{center}
\includegraphics[width=0.7\textwidth]{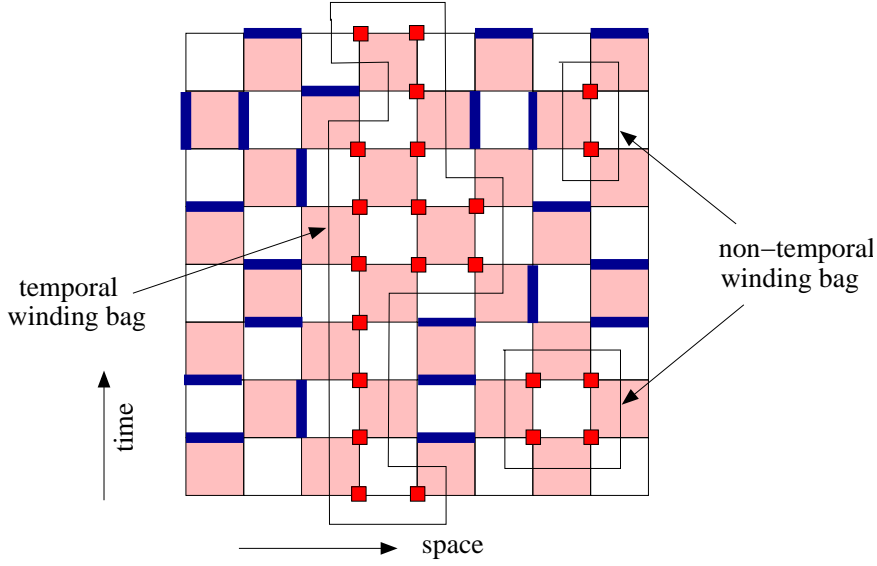}
\end{center}
\caption{\label{fig2} An illustration of a ``fermion-bag'' configuration with one temporal winding bag and two non-temporal winding bags.}
\end{figure}

The fermion bag approach also suffers from a sign problem in the presence of a chemical potential. But the sign problem tracks the physics of the model very closely. To see this note that the partition function is again given by Eq.~(\ref{fbpf}) except that now $W[n]$ is the free Dirac operator operator with the chemical potential given in Eq.~(\ref{Dmu}) where the sites connected to $n_{x,\mu}=1$ are dropped. Again it is no longer possible to argue that $\mathrm{Det}(W[n]) \geq 0$ when $\mu \neq 0$. However, this sign problem is qualitatively different. Since the chemical potential only enters through fermion world lines that wrap around the temporal direction, the chemical potential completely drops out of the determinant of a fermion bag which lives completely inside the space time volume. We call these non-temporal winding bags and two such bags are shown in Fig.~\ref{fig2}. The fermions hopping within this bag will never have a fermion world line with a non-zero temporal winding. Only bags with a non-zero temporal winding are sensitive to the chemical potential. One such bag is also shown in Fig.~\ref{fig2}. At large $U$ and small temperatures (large temporal direction), such bags are exponentially suppressed. Thus, the sign problem is naturally absent for small chemical potentials and low temperatures at large $U$ in the fermion bag approach as dictated by physics.

For small $U$ when the fermions are massless, temporal winding bags proliferate and the fermion bag approach also suffers from a severe sign problem in the presence of a chemical potential. We do claim to have a solution to this sign problem in the case of a single flavor of staggered fermion. However, in the next section we argue that the fermion bag approach allows us to solve this sign problem with an even number of flavors.

\section{New Solutions to Sign Problems}

The fermion bag approach also offers new solutions to some seemingly unsolvable sign problems. In order to appreciate this 
consider the action of the $N$ flavor model given by the action
\begin{equation}S = - \sum_{x,y} {\overline\psi}_x \ D(\mu)_{x,y} \ \psi_y\ + U \ \sum_{x,\alpha} ({\overline\psi}_{x+\hat\alpha}\psi_x)\ ({\overline\psi}_x\psi_{x+\hat\alpha})
\label{actN}
\end{equation}
where $\psi_x$ is an $N$-component column vector and ${\overline\psi}_x$ is an N-component row vector. This action is invariant under a $U(N)\times U(N)$ symmetry. The partition function in the auxiliary field method turns out to be
\begin{equation}
Z = \int [d\phi] \ \Bigg\{\mathrm{Det}(M([\phi]))\Bigg\}^N
\label{convpfN}
\end{equation}
where the matrix $M[\phi]$ is the same as the one-flavor model given in Eq.~(\ref{muM}). Since the $\mathrm{Det}(M[\phi])$ is a general complex number in the presence of a chemical potential, its $N$th power remains complex for all $N$. Unfortunately, this sign problem remains unsolved within the fermion bag approach as well. On the other hand consider the model given by the action
\begin{equation}
S = - \sum_{x,y} {\overline\psi}_x \ D(\mu)_{x,y} \ \psi_y
\ - \frac{U}{(N!)^2} \ \sum_{x,\alpha} 
\Bigg\{{(\overline\psi}_{x+\hat\alpha}\psi_x)\ 
({\overline\psi}_x\psi_{x+\hat\alpha})\Bigg\}^N.
\label{modactN}
\end{equation}
This action is again invariant under the same $U(N)\times U(N)$ chiral symmetry. In the auxiliary field method one will need many auxiliary fields to convert the $4N$-fermion term to a bi-linear. Further, even with these additional fields, it is difficult to see why the determinant of the fermion matrix that will arise will be positive for any value of $N$ for the same reasons outlined above. On the other hand in the fermion bag approach this modified model is described by the partition function
\begin{equation}
Z = \sum_{n_{x,\alpha}=0,1} \Big(\prod_{x,\alpha} U^{n_{x,\alpha}}\Big) 
\Bigg\{\mathrm{Det}(W[n])\Bigg\}^N.
\label{fbpfN}
\end{equation}
Since $\mathrm{Det}(W[n])$ is real, there is no sign problem with even $N$. Thus, the fermion bag approach is able to solve a sign problem that seems unsolvable with the auxiliary field method. In this context we must point out that there are indeed other actions that are invariant under $U(N)\times U(N)$ symmetry whose partition functions can be written without a sign problem using the auxiliary field method for even values of $N$.

\section{The Monte Carlo Method}
\label{MC}

In order to solve the massless Thirring model using the fermion bag approach, in this section we discuss two update algorithms: (1) Local Heal Bath, and (2) Directed Path algorithm. For generality we discuss these algorithms for any dimension $d$, although the current work is focused on $d=3$. We will argue that these two update algorithms together provide an efficient way to solve the problem for $U \gtrsim 1.2$. When $0.2 < U < 1.2$ these algorithms do slow down dramatically, however they continue to provide a useful way to solve the problem. The efficiency may be comparable if not superior to the HMC method. For values of $U < 0.2$ the HMC algorithm will be a better approach. However, it is likely that the diagrammatic Monte Carlo provides a better algorithm for small and intermediate values of $U$ \cite{prokof'ev:250201}.

The configurations are described by $n_{x,\alpha} =0,1$ bond variables. A dimer is represented by $n_{x,\alpha} = 1$. We will assume that $\alpha$ can take any of the $2d$ values: $\alpha = \pm 1, \pm 2, ...\pm d$, where the negative signs indicate negative directions. This means $n_{x,\alpha} \equiv n_{x+\hat{\alpha},-\alpha}$. For convenience we also define site variables $m_x=0,1$. A monomer is represented by $m_x=1$. To begin with we set $m_x=0$ at all sites. It is useful to remember that a site $x$ that belongs to a fermion bag should have both $m_x=0$ and $n_{x,\alpha} = 0, \forall \alpha$.  The parity of a site $x$ is defined as $\varepsilon_x = (-1)^{x_1+x_2+...+x_d}$.

\subsection{Local Heat Bath}

The first update we discuss creates and destroys dimers. This is accomplished with a local heat update. The exact update is as follows:
\begin{enumerate}
\item Pick a site $x$ at random on the lattice.
\item There are $2d+1$ possible values for $\{n_{x,\alpha}\}$: $n_{x,\alpha} = 1$ for one of the $2d$ values of $\alpha$ or $n_{x,\alpha} = 0, \forall \alpha$. In this latter configuration let us label the fermion bag that is connected to the site $x$ as ${\cal B}_x$.  Let $W[{\cal B}_x]$ be the free Dirac matrix inside this bag. If $\mathrm{Det}(W[{\cal B}_x]) = 0$, the update ends without changing the original configuration. Otherwise the update proceeds to the next step.
\item Let $\omega_\alpha = U |((W[{\cal B}_x])^{-1})_{x,x+\hat{\alpha}}|^2$ for the $2d$ values of $\alpha$. We set $\omega_0=1$.
\item We pick $\alpha$ with probability 
\begin{equation}
P_\alpha = \frac{\omega_\alpha}{\sum_\alpha \omega_\alpha}
\end{equation}
\item If $\alpha=0$ we set $n_{x,\alpha} = 0$ for all values of $\alpha$ and stop. Otherwise we set $n_{x+\alpha} = 1$ and others to zero then stop.
\end{enumerate}
We define a sweep as consisting of $(L/2)^3$ local heat updates updates.

The most time consuming step of this local heat bath update is the computation of $(W^{-1}[{\cal B}_x])_{x,x+\hat{\alpha}}$. It clearly depends on the size of the bag ${\cal B}_x$. When the typical bag size does not scale with the volume the time to compute the inverse also does not scale with the volume. This is the reason for the efficiency of the Monte Carlo method in the fermion bag method. We will show that this is indeed the case when $U \gtrsim 1.0$.

In order to compute the inverse we set the vector $b_y=\delta_{x,y}$ and then solve the equation $W v = b$. Practically we solve $(-W^2) v = (-W b)$ and since $(-W^2)$ is a positive definite matrix, we can use the conjugate gradient method. The convergence of the answer is checked by the parameter $\gamma = |W v - b|^2$. If this norm is less than $10^{-20}$ we assume that the solution has been found. Another useful norm is $\gamma' = |(-W^2) v + W b|^2$ and can be used to detect exact zero modes of $W$ using the conjugate gradient method. Note that $(-W b)$ eliminates the zero mode subspace from the source vector and the space on which conjugate gradient acts. Thus the conjugate gradient method can always make $\gamma'$ arbitrarily small. If $\gamma$ cannot be made smaller than $10^{-20}$ even when $\gamma' < 10^{-30}$, we declare that configuration to have an exact zero mode. This method appears to work reliably.

\subsection{Directed Path Update}

The second update preserves the number of dimers but moves them around. This update is similar to the directed path update discussed in \cite{Adams:2003cc} and reduces to it in the limit of large $U$. The philosophy behind it is similar to the worm algorithm \cite{Prokof'ev:2001zz}. The update is as follows:
\begin{enumerate}
\item Pick a site $x$ at random. 
\item If $n_{x,\alpha} = 0, \forall \alpha$ the update stops. If not we label $x$ and all sites with the same parity as active sites. The sites with the opposite parity are labeled as passive sites. We then perform either an active or a passive update depending on our current site as discussed below. After each update we move through the lattice according to the rules of the update until we return to the first site, where the update ends.
\item[] {\bf Active Update}: If we are on an active site $x$, we do one of four things depending on the configuration on the site.
\begin{enumerate}
\item If $x$ is the first site such that $m_x=0$ and $n_{x,\alpha}=1$, then we set $n_{x,\alpha}=0$ and $m_x=1$ and $m_{x+\hat{\alpha}}=1$. In other words we break a dimer into two monomers. The update then moves to the site $x+\hat{\alpha}$.
\item If $x$ is not the first site such that $n_{x,\alpha}=1$ and we just came to the site from the previous site $x+\hat{\beta}$, then we set $n_{x,\beta} = 1$, $m_{x+\beta}=0$ and $m_{x+\alpha} = 1$. The update then moves to the site $x+\hat{\alpha}$.
\item If $x$ is not the first site such that $n_{x,\alpha} = 0$ for all the values of $\alpha$ and we just came to the site from the previous site $x+\hat{\beta}$, then we pick a direction $\gamma$ with probability $P_\gamma(x)$ to be discussed below. We set $m_{x+\hat{\beta}}=0$ and $m_{x+\hat{\gamma}}=1$. In other words we move the monomer from the site $x+\hat{\beta}$ to $x+\hat{\gamma}$.
\item If $x$ is the first site such that $m_x=1$ and $n_{x,\alpha}=0$, then we would have returned to it from the neighboring site $x+\hat{\beta}$ such that $m_{x+\hat{\beta}}=1$. We then set $m_x=0$, $m_{x+\hat{\beta}}=0$ and $n_{x,\beta}=1$. The update then ends.
\end{enumerate}
\item[] {\bf Passive Update}: If we are on a passive site $x$ then we must have $m_x=1$. We pick one of the $2d+1$ directions $\alpha$ including $0$ at random. If $\alpha = 0$ the update remains on the same site, we get a contribution to the two-point correlation function discussed below. If $\alpha \neq 0$ the update moves to the neighboring active site $x+\hat{\alpha}$.
\end{enumerate}
Let us now discuss the probability $P_\gamma(x)$ on an active site $x$ such that $n_{x,\alpha} = 0$ for all values of $\alpha$ and such that $m_y=1$ where $y=x+\hat{\beta}$. The site $x$ is associated to a fermion bag say ${\cal B}_x$. Note that the passive site $y$ is not in the bag. Let $x0$ be another active site which is not in the bag, but contains a neighboring site which is in the bag. Thus, both $x0$ and $y$ ``touch'' the bag ${\cal B}_x$ but are not a part of it. Let us extend the bag to include both $y$ and $x0$ call the extended bag ${\cal B}_{x,x0,y}$. The probability $P_\gamma$ is then given by
\begin{equation}
P_\gamma(x) = \frac{\omega_{x+\hat{\gamma}}}{\sum_\gamma \omega_{x+\hat{\gamma}}}
\end{equation}
where $\omega_z = |[(W[{\cal B}_{x,x0,y})^{-1}]_{x_0,z}|^2$. It is possible to show that while $\omega_z$ depends on $x_0$, $P_\gamma$ does not. Further, if $x+\hat{\gamma}$ does not belong to the bag $B_{x,x0,y}$ $P_\gamma(x) = 0$.

The most time consuming step in this update is the computation of the probabilities $P_\gamma(x)$ on an active site $x$. Fortunately, as long as the fermion bags are not disturbed $\omega_y$ does not change on any site $y$ inside the bag. So $P_\gamma(x)$ is then simple to compute. However, if the fermion bag is disturbed the extra effort in computing the inverse is necessary. For large values of the $U$ the bags are small and the effort does not grow with the volume.

During the passive update on there is a probability to remain on the same site. This can be shown to be the correct probability to create a monomer at that site along with another monomer at the first site. Hence it contributes to the two point correlation function
\begin{equation}
G(x,y) = \Big\langle \overline{\psi}_x\psi_x \overline{\psi}_y\psi_y\Big\rangle
\end{equation}
Thus, we can compute this correlation function during this update. Here we use this to compute the susceptibility. We have tested the algorithm against exact calculations on a small lattice and the results are given in the appendix.

\section{Distribution of Fermion Bags}

Fermion bags encode the fermionic physics, understanding their properties is an important research problem in itself. For example the eigenvalue distribution of the corresponding Dirac operator could be interesting. What role do the low eigenvalues play? Can their distribution be described by some simplified theory like random matrix theory? However, we postpone such studies to the future. Here we focus on computing a much simpler quantity, namely the size distribution of the fermion bags as a function of the coupling $U$. This quantity helps us understand the efficiency of the fermion bag approach and identify the range of $U$ where the approach is clearly superior.

\begin{figure}[t]
\begin{center}
\includegraphics[width=0.7\textwidth]{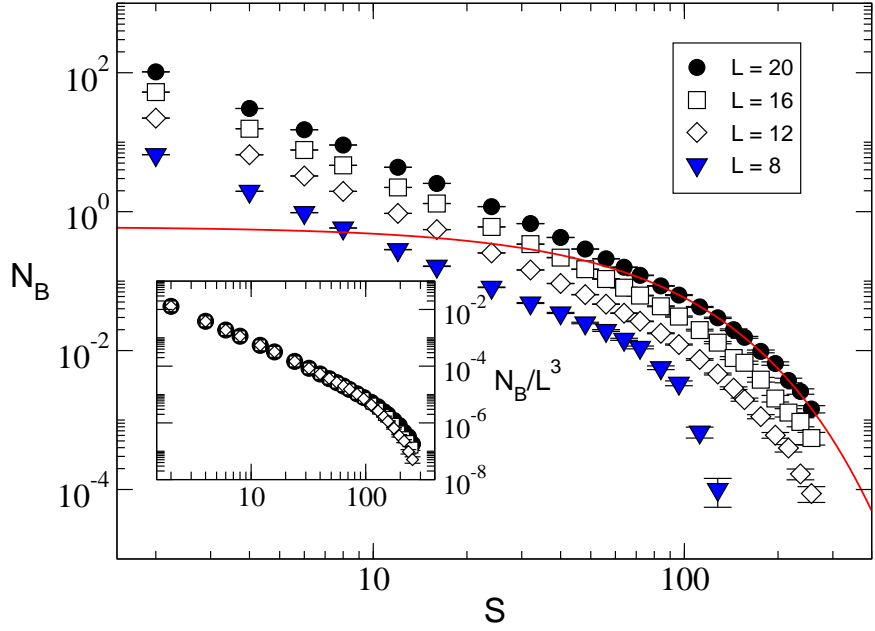}
\end{center}
\caption{\label{fig3} Plot of the average number of fermion bags $\langle N_B \rangle$ of size $S$ as a function of the $S$ for $L=8,12,16$ and $20$ at $U=1.3$. The solid line is an exponential fit as discussed in the text. The inset shows the same plot ($L=8$ data has been omitted) scaled with the volume. The data collapse shows that the density of fermion bags of a given size does not depend on the volume.}
\end{figure}

\begin{figure}[t]
\begin{center}
\includegraphics[width=0.7\textwidth]{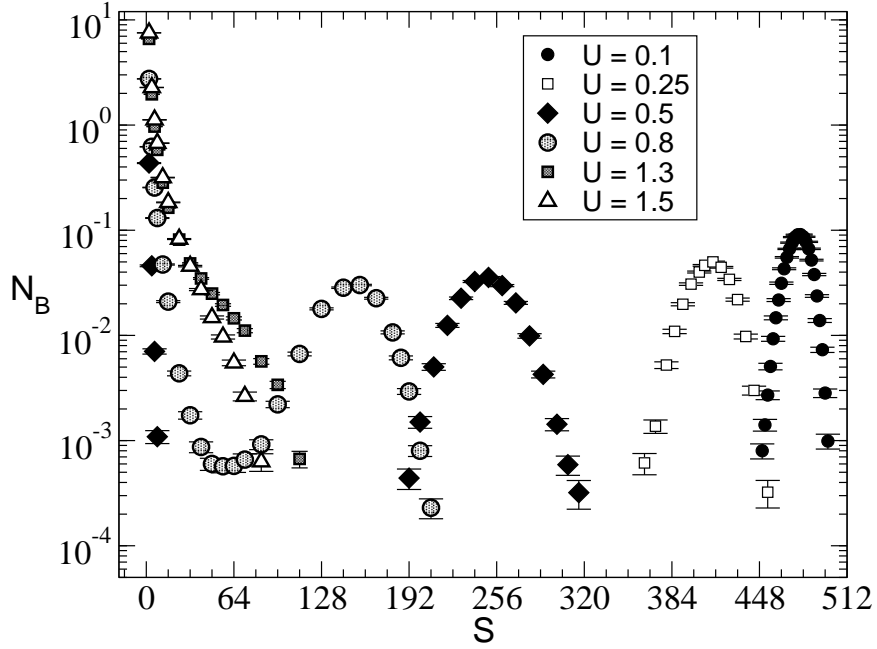}
\end{center}
\caption{\label{fig4} Plot of the average distribution of fermion bags $\langle N_B \rangle$ of size $S$ as a function of the $S$ for $U=0.1,0.25,0.5,0.8,1.3$ and $1.5$ for $L=8$. We note that the distribution changes qualitatively between large and small $U$. }
\end{figure}

Let $N_B(S)$ be the number of bags of size $S$ in a single configuration. In Fig.~\ref{fig2} we plot the average of $N_B$ over the ensemble of configurations generated by the algorithm at $U=1.3$ for $L=8,12,16$ and $20$. The figure shows that the number of bags of a given size increases with the volume but the density of the bags of a given size remains constant. Indeed the three data points for $L=12,16,20$ collapse on a single curve once the density is plotted as shown in the inset of the figure. We find that $N_B(S)$ drops like a power for small values of $S$, but somewhere around $S \gtrsim 100$, a sudden drop in $N_B$ is observed. This behavior is similar for all values of $L$ except that the sudden drop moves slightly. This we attribute to a finite size effect. For large values of $S$, we find $N_B(S)$ decays exponentially as a function of $S$. Assuming that the bag size represents a three dimensional lattice volume then we naturally expect
\begin{equation}
N_B(S) = A\ \exp(-M^3 S)
\end{equation}
where $M$ is a lattice mass scale. The data for $L=20$ fits well to this form when $S \geq 84$, we get $A = 0.61(2)$, $M=0.286(1)$ and a $\chi^2/DOF = 0.8$. This fit is shown as a solid line in the plot of Fig.~\ref{fig3}. It is tempting to relate the scale $M$ with the mass of the fermion. However, given that the critical point where the fermion becomes massless is roughly around $U\sim 0.25$ \cite{Debbio:1997dv} we think that the scale $M$ is not the mass of the fermion. We postpone the study of its origin to a later publication. 

\begin{figure}[t]
\begin{center}
\includegraphics[width=0.7\textwidth]{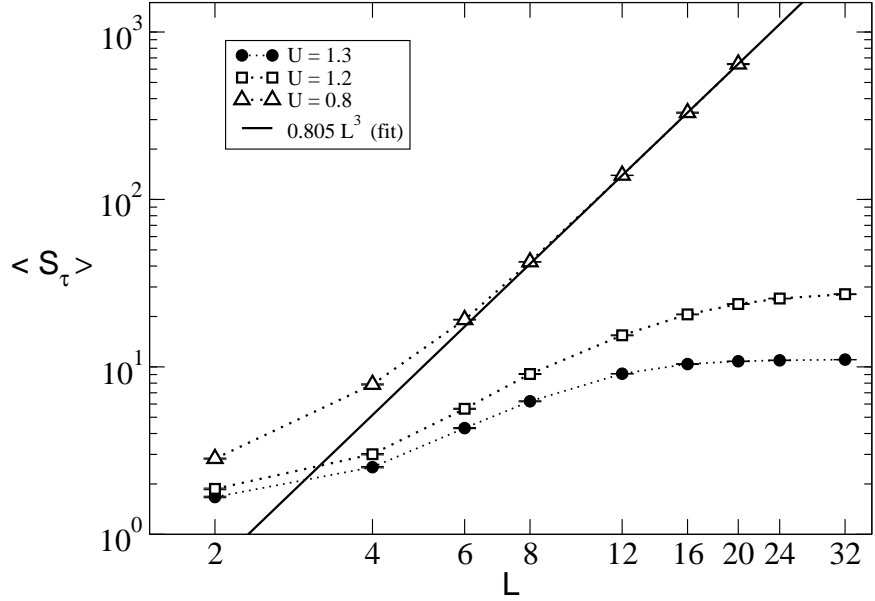}
\end{center}
\caption{\label{fig5} Plot of the typical bag size as a function of $L$ for three different values of $U$. The solid line is a fit to the form $A L^3$. Note the $L$ axis is shown on a logarithmic scale.}
\end{figure}

In Fig.~\ref{fig4} the bag distribution is shown for many values of $U$ at $L=8$. We see that distribution changes qualitatively when $U$ becomes small. Instead of decaying exponentially with size, a bump develops in the bag distribution at a value of $S$ of about half the system size. The position of the bump then begins to grow till it becomes as big as the system size for very small $U$. This is easily understandable. As $U$ reduces the bags merge so that now the whole lattice becomes one large bag with small regions of confined (or paired) fermions which in a sense form ``dual bags''.

It is useful to define a typical size of a fermion bag as the size of the bag that one encounters on an average during the update. More precisely we pick a site at random and define $S_\tau$ as the size of the bag associated to that site. We can then average it over the ensemble of the configurations generated. It is easy to argue that
\begin{equation}
\langle S_\tau \rangle = \frac{1}{L^3} \Bigg\langle \Big(\sum_B S_B^2\Big) \Bigg\rangle
\end{equation}
where $S_B$ is the size of the bag $B$ and the sum is over all the bags in a given configuration. In Fig.~\ref{fig5} we plot $\langle S_\tau \rangle$ as a function of $L$ for $U=1.3, 1.2$ and $0.8$. The solid line is the fit given by $0.0806(2) L^3$. The figure shows clearly that for $U = 1.2$ and $1.3$ the typical fermion bag size begins to saturate, indicating that the Dirac matrix used in the conjugate gradient has a typical size independent of the lattice size for large volumes. When $U=0.8$ this advantage is no longer valid since the bags begin to grow with the spatial volume. Thus, the fermion bag approach is guaranteed to be efficient only when $U \gtrsim 1.2$. For smaller $U$ the bag approach continues to be an alternative approach but becomes less attractive. However note that at $U=0.8$ the bags only occupy roughly $1/10$ the size of the system. Thus, the bag approach may continue to be competitive at intermediate values of $U$ and moderate values of $L$.

\section{Results in the Massless Thirring Model}

In this section we present some results obtained using the fermion bag approach in the massless Thirring model with one flavor of staggered fermions in three dimensions. We focus on the following three observables:
\begin{enumerate}
\item Chiral condensate susceptibility: 
\begin{equation}
\chi = \frac{U}{L^3}
\sum_{x,y} \ 
\Big\langle \overline{\psi}_x\psi_x \ \overline{\psi}_y\psi_y \Big\rangle.
\end{equation}
Here the factor $U$ ensures that in the $U = \infty$ limit the chiral condensate $\Sigma$ is finite.
\item Fermion charge susceptibility:
\begin{equation}
\langle Q_f^2 \rangle = 
\sum_{x\in S,y\in S'} \ \Big\langle J^f_{\alpha,x} J^f_{\alpha, y} \Big\rangle.
\end{equation}
Here $S$ and $S'$ are two different two-dimensional surfaces perpendicular to the direction $\alpha$ and
\begin{equation}
J^f_{\alpha,x} = 
\frac{\eta_{x,\alpha}}{2}\Big\{\psib_x \psi_{x+\alpha} + \psib_{x+\alpha}\psi_x\Big\}
\end{equation}
is the conserved fermion current. In the bag formulation one can show that
\begin{equation}
\langle Q^2_f\rangle = \Bigg\langle
\frac{1}{2}\sum_{x\in S,y\in S'} \eta_{x,\alpha}  \eta_{x,\alpha}  
\Big[(D^{-1})_{x,y+\alpha}(D^{-1})_{x+\alpha,y} + (D^{-1})_{x,y}(D^{-1})_{x+\alpha,y+\alpha}\Big]\Bigg\rangle
\end{equation}
where $D^{-1}$ is the inverse of the free Dirac operator inside the bag containing all the four points $x,y,x+\alpha,y+\alpha$. If any of these points is not part of the bag then the corresponding contribution is set to zero. It is easy to check that $Q_f^2$ defined here is independent of the surfaces chosen on every bag configuration.
\item Chiral charge susceptibility:
\begin{equation}
\langle Q_\chi^2 \rangle = \sum_{x\in S,y\in S'} 
\ \Big\langle J^\chi_{\alpha,x} J^\chi_{\alpha, y} \Big\rangle
\end{equation}
where the notation is the same as for the fermion charge susceptibility, except the conserved chiral current is given by
\begin{equation}
J^\chi_{\alpha,x} = 
\frac{\varepsilon_x\eta_{x,\alpha}}{2}\Big\{\psib_x \psi_{x+\alpha} - \psib_{x+\alpha}\psi_x\Big\}.
\end{equation}
On every bag configuration let us define
\begin{equation}
q_\chi = \sum_{x\in S(B)} \varepsilon_x\eta_{x,\alpha} (D^{-1})_{x,x+\alpha} \ +\  
\sum_{x\in S(C)} 2 \varepsilon_x\ \,
\end{equation}
where $S(B)$ is the set of sites on the two-dimensional surface $S$ which belongs to some fermion bag while $S(C)$ are the remaining set of sites on the surface. For a given bag configuration we find that $q_\chi$ is also independent of the surface. Further
\begin{equation}
\langle Q_\chi^2 \rangle = \langle Q_f^2\rangle + \langle q_\chi^2 \rangle,
\end{equation}
where the average is over the fermion bag configurations generated.
\end{enumerate}

For large values of $U$ chiral symmetry is broken spontaneously and fermions acquire a mass. Chiral perturbation theory predicts that \cite{Hasenfratz:1989pk},
\begin{subequations}
\label{eq:largeu}
\begin{eqnarray}
\langle q_\chi^2 \rangle &=& 4 \rho_s L [1 + \frac{0.224}{\rho_s L} + \frac{a}{(\rho_s L)^2}] \\
\chi &=& \frac{L^3\Sigma^2}{2} \Bigg[ 1 + \frac{0.224}{\rho_s L} + \frac{b}{(\rho_s L)^2}\Bigg]
\end{eqnarray}
\end{subequations}
and that  $\langle Q_f^2 \rangle$ goes to zero exponentially. The parameters $\rho_s$, $\Sigma$, $a$ and $b$ are the low energy constants and can be determined from fitting the data. The constant $\rho_s$ is a mass scale and plays the role of $F_\pi$ in four dimensional chiral perturbation theory. The chiral condensate in the chiral limit is given by $\Sigma$. The factor $4$ in the expression for $\langle q^2_\chi \rangle$ is not standard. However in our definition the charge $q_\chi$ takes only even values at $U=\infty$ and so $\rho_s$ will not be properly normalized without this factor. For small values of $U$ chiral symmetry is restored, but due to the presence of massless fermions we expect
\begin{subequations}
\label{eq:smallu}
\begin{eqnarray}
\chi &=& U (\chi_0 + \chi_1/L + \chi_2/L^2 + ...) \\
\langle q_\chi^2 \rangle &=& q_1/L + q_2/L^2 + q_3/L^3.... \\
\langle Q_f^2 \rangle &=& \gamma_0 + \gamma_1/L + \gamma_2/L^2+...
\end{eqnarray}
\end{subequations}
For free staggered fermions (i.e., when $U=0$ ) we find $\chi_0 \approx 1.01093$ and $\gamma_0 \approx 0.37085$ while $\langle q_\chi^2 \rangle = 0$. 

\begin{figure}[t]
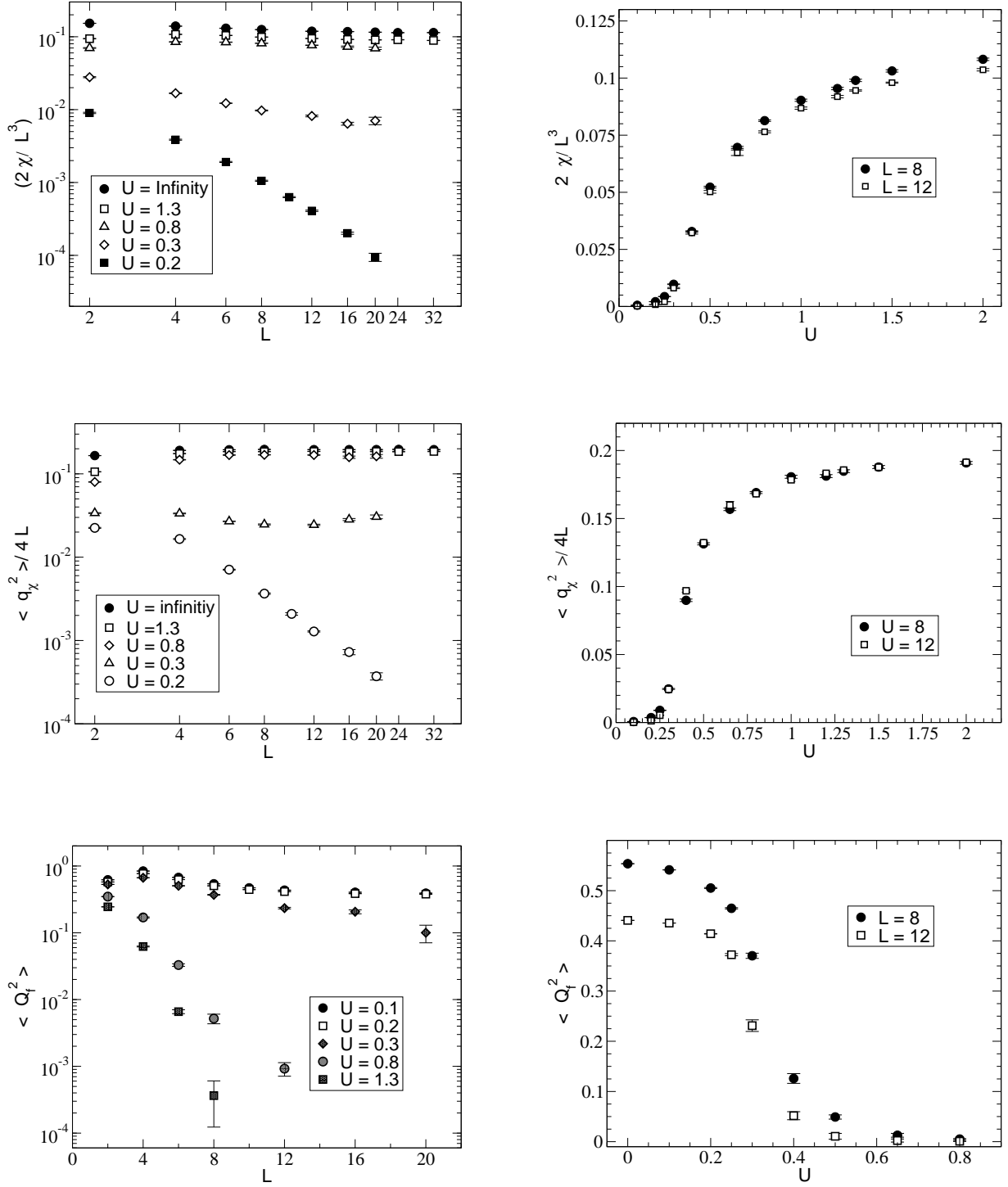

\begin{center}
\vbox{
\hbox{
\includegraphics[width=0.45\textwidth]{fig6.eps}
\hskip0.5in
\includegraphics[width=0.45\textwidth]{fig7.eps}
}
\vskip0.5in
\hbox{
\includegraphics[width=0.45\textwidth]{fig8.eps}
\hskip0.5in
\includegraphics[width=0.45\textwidth]{fig9.eps}
}
\vskip0.5in
\hbox{
\includegraphics[width=0.45\textwidth]{fig10.eps}
\hskip0.5in
\includegraphics[width=0.45\textwidth]{fig11.eps}
}
}
\end{center}
\caption{\label{fig6} Plots of $2 \chi/L^3$, $\langle q_\chi^2\rangle/4 L$ and $\langle Q_f^2\rangle$ as a function of the lattice size for various values of $U$ is shown in the left figures. The same quantity is plotted as a function of $U$ for $L=8,12$ on the right.}
\end{figure}

Our data are consistent with these expectations qualitatively for both large and small values of $U$. In Fig.~\ref{fig6} we plot the three observables scaled appropriately as a function of $L$ for different values of $U$ (left plots) and for $L=8,12$ as a function of $U$ (right plots). We see that there the finite size scaling changes abruptly between $U=0.2$ (symmetric phase) and $U=0.3$ (broken phase). In particular $\Sigma$ and $\rho_s$ are non-zero for $U \geq 0.3$, but vanish for $U \leq 0.2$. On the other hand $\langle Q_f^2 \rangle$ remains non-zero when $U\leq 0.2$ and begins to drop significantly when $U \geq 0.3$. Thus, there is a clear phase transition in the model somewhere between these two couplings. We will study this quantum phase transition quantitatively in the next section. 

\begin{figure}[t]
\begin{center}
\vbox{
\includegraphics[width=0.45\textwidth]{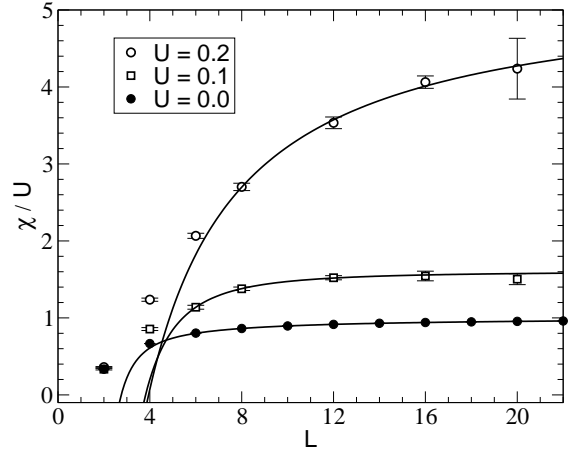}
\vskip0.5in
\includegraphics[width=0.45\textwidth]{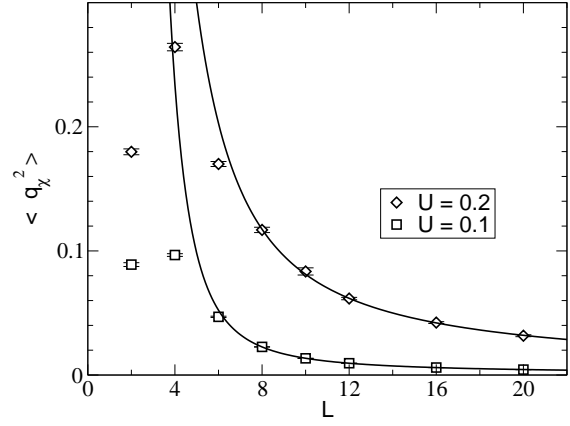}
\vskip0.5in
\includegraphics[width=0.45\textwidth]{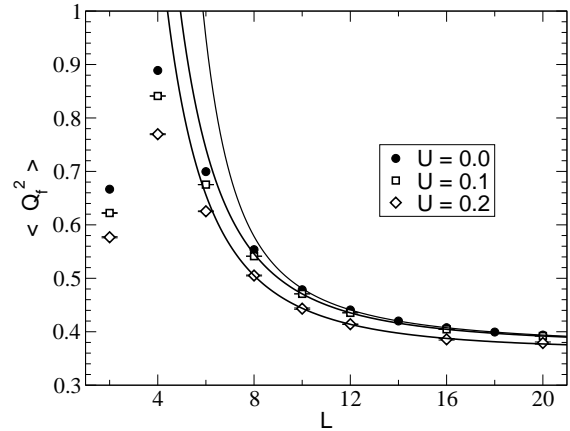}
}
\end{center}
\caption{\label{fig7} Plots of $\chi$, $\langle q_\chi^2\rangle $ and $\langle Q_f^2\rangle$ as a function of the lattice size for small values of $U$. The solid lines are fits to the expected forms in}
\end{figure}

Quantitatively, we can fit our data to expectations from chiral perturbation theory only for $U \geq 1.2$ where the fermion bag approach allows us to go to large volumes. Typically we have found that the fits to the finite size scaling forms given in Eq.~(\ref{eq:largeu}) are good in the range $16 \leq L \leq 32$. The values of the low energy constants that give good fits for $U \geq 1.2$ are given in table \ref{tab:largeU}. We observe that a change from $U=\infty$ to $U=1.2$ leads to only about $7\%$ change in the mass scale $\rho_s$ and about $10\%$ change in the chiral condensate. This again shows that the effort of conventional determinantal methods is unnecessary and the fermion bag approach is the better method in this range of couplings. For small values of $U$, in the symmetric phase, the fermion bag approach slows down considerably. Hence we have been able to perform calculations only up to $L=20$. In Fig.~\ref{fig7} we show results for $U=0.0,0.1$ and $U=0.2$. In table \ref{tab:smallu} we show the results of the fits to Eq.~(\ref{eq:smallu}).

\clearpage

\clearpage

\begin{table}[t]
\begin{center}
\begin{tabular}{c||c|c|c||c|c}
\hline
\hline
$U$ & $\rho_s$ & $a$ & $\chi^2$/DOF & $\Sigma$ & $\chi^2$/DOF\\
\hline
$\infty$ & $0.191(1)$ & $-0.48(5)$ & $0.4$ & $0.3302(3)$ &  $1.7$\\
$1.5$ & $0.184(1)$ & $-0.52(7)$ & $0.8$ & $0.2980(3)$ & $0.7$ \\
$1.3$ & $0.181(1)$ & $-0.56(10)$ & $0.3$ & $0.2927(3)$ & $0.9$ \\
$1.2$ & $0.176(2)$ & $-0.23(11)$ & $0.9$ &$0.2881(3)$ & $1.9$ \\
\hline
\hline
\end{tabular}
\end{center}
\caption{\label{tab:largeU} The coefficients in Eq.~(\ref{eq:largeu}) obtained by fitting the data. In the fits of $\chi$ the coefficient $b$ was set to zero and $\rho_s$ was fixed from the second column. The data in the range $16 \leq L \leq 32$ were used for fits of $\langle q_\chi^2 \rangle$ while the range $8 \leq L \leq 32$ was used for $\chi$.}
\vskip0.2in
\begin{center}
\begin{tabular}{c||c|c|c|c|c}
\hline
U   & $\chi_0$    & $\chi_1$    & $\chi_3$ & $\chi_5$ & $\chi^2$/DOF \\
\hline
0.0 & 1.010930(1) & -1.11288(5) & -3.61(1) & 69(1)   & 0.6           \\
0.1 & 1.55(11)    & -0(2)       & -95(32)  & --      & 0.3          \\
0.2 & 5.35(14)      & -21(1)      & --       & --      & 0.3           \\
\hline
U   & $q_1$    & $q_3$    & $q_5$ & $q_7$ & $\chi^2$/DOF \\
\hline
0.1 & 0.077(3) & 4.3(5) & 150(27) & - & 0.6 \\
0.2 & 0.58(2) & 23(2) & - & - & 0.35 \\
\hline
U   & $\gamma_0$    & $\gamma_2$    & $\gamma_4$ & $\gamma_6$ & $\chi^2$/DOF \\
\hline
0.0 & 0.370840(1) & 8.858(2) & 92.8(7) & 12204(100) & 0.9 \\
0.1 & 0.3734(2) & 6.45(6) & 425(6) & -9558(133) & 0.5 \\
0.2 & 0.365(2) & 4.3(5) & 445(41) & -9493(960) & 2.0 \\ 
\hline
\end{tabular}
\caption{\label{tab:smallu} The coefficients in Eq.~(\ref{eq:smallu}) obtained by fitting the data. The missing coefficients have been assumed to be zero in the fit. For $U=0.0$ we computed the coefficients exactly but assigned an error of $10^{-6}$ uniformly. The results were then fit for $L > 10$. For $U=0.1$ and $0.2$ data from $8 \leq L \leq 20 $ were used in the fit.}
\end{center}
\end{table}

While our data fits to the expected finite size scaling forms even for small $U$, due to the small system sizes used in the calculations we do not claim to be able to extract the $L$ dependence reliably except for the free theory. In particular we expect large systematic errors in the determination of the coefficients shown in table \ref{tab:smallu}. In fact we have used the free theory to guide our fits. For example it is tempting to associate $\gamma_0 = 0.370840..$ to a universal constant for free massless fermions. It is likely that this does not change with $U$ since we expect that the infrared physics to be that of free massless fermions. The small change that we observe in our fits perhaps is due to systematic errors associated to fitting the data at smaller lattice sizes. Further work is clearly necessary to reliably understand the small $U$ regime.

\section{Quantum Phase Transition}

We next focus on the quantum phase transition in the model. This transition has already been studied earlier using mean field theory \cite{PhysRevLett.59.14}, auxiliary field method \cite{DelDebbio:1995zc} and through a formulation as a strongly coupled $U(1)$ lattice gauge theory with scalar fields \cite{PhysRevLett.59.14,Barbour:1998yc}. In the latter two studies, algorithmic difficulties forced the use of a non-zero fermion mass. This usually makes the study of finite size scaling close to a second order transition more tricky since the fermion mass introduces a new length scale. However, by making a judicious choice of the scaling relations, both these studies concluded that the transition was consistent with a second order transition. The critical coupling was estimated to be $U_c=0.25(1)$ \cite{DelDebbio:1995zc} and $U_c=0.28(1)$\cite{Barbour:1998yc}, while the critical exponents were determined to be $\delta \approx 2.5(4)$, $\beta \approx 0.71(9)$ \cite{DelDebbio:1995zc} and $\delta \approx 3.1(3)$, $\nu \approx 0.88(9)$ \cite{Barbour:1998yc}. For comparison, mean field theory in $d=3$ predicts $U_c=0.177(6)$, $\delta=2$ and $\beta=1$ \cite{PhysRevLett.59.14,Moshe:2003xn}.

The errors we quote here include systematic errors estimated from the different fitting procedures used in the previous work. In the present work we estimate these parameters again in the fermion bag approach. 

\begin{figure*}[t]
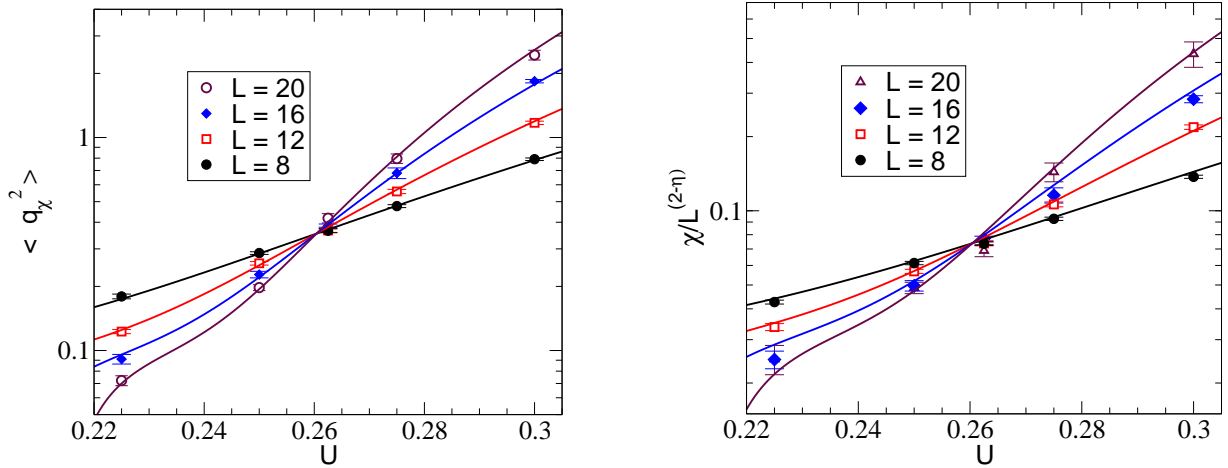

\begin{center}
\hbox{
\includegraphics[width=0.45\textwidth]{fig15.eps}
\hskip0.5in
\includegraphics[width=0.45\textwidth]{fig16.eps}
}
\end{center}
\caption{\label{fig8} Plots of $\langle q_\chi^2\rangle$  and $\chi/L^{2-\eta}$ as a function of $U$ for $L=8,12,16$ and $20$. The solid lines show the combined fit of all the data to Eq.~(\ref{critfunc}) as discussed in the text. Based on the fits we find the critical point to be $U=0.2604(5)$, $\nu = 0.87(2)$ and $\eta=0.62(2)$}
\end{figure*}

One main difference as compared to the previous work is that we work with exactly massless fermions which helps with a clean finite size scaling analysis. Assuming the transition to be a conventional second order phase transition we expect
\begin{subequations}
\label{critfunc}
\begin{eqnarray}
\langle q_\chi^2 \rangle &=& \kappa_0 + \kappa_1 (U - U_c) L^\frac{1}{\nu} + 
\kappa_2 (U - U_c)^2 L^\frac{2}{\nu} + ... 
\\
\chi/L^{2-\eta} &=&  f_0 + f_1 (U - U_c) L^\frac{1}{\nu} + 
f_2 (U - U_c)^2 L^\frac{2}{\nu} + ...
\end{eqnarray}
\end{subequations}
A combined fit of our data to these relations give the following results: 
\begin{center}
\begin{tabular}{c|c|c|c|c|c|c|c|c|c|c}
\hline \hline
$\eta$ & $\nu$ & $U_c$ & 
$f_0$ & $f_1$ & $f_2$ & $f_3$ & 
$\kappa_0$ & $\kappa_1$ & $\kappa_2$ & $\kappa_3$ \\
0.62(2) & 0.87(2) & 0.2604(5) & 0.074(4) & 0.11(1) & 0.11(2) & 0.04(1) &
0.354(6) & 0.68(4) & 0.65(9) & 0.24(5) \\
\hline \hline
\end{tabular}
\end{center}
with a $\chi^2/$DOF of 1.6. The data used in the combined fit and the fits themselves are shown in Fig.~\ref{fig8}. Using hyper-scaling relations $2 \beta = \nu (d - 2 + \eta)$ and $\delta = (d + 2 - \eta)/(d - 2 + \eta)$ we estimate $\beta \approx 0.71(4)$ and $\delta \approx 2.70(4)$. While, these results are in complete agreement with the earlier work, our values seem more accurate. One obvious caveat is that our results are also obtained on rather small lattices compared to what is available in bosonic models. Thus, we may be underestimating some systematic errors. On the other hand we have exactly massless fermions so some of the fitting procedures are cleaner. In any case accurate results on larger lattices are desirable to confirm these findings.

It was pointed out in \cite{PhysRevD.51.5816} that the Thirring model is different from the Gross-Nevue (GN) model in many ways. However, recent work suggests that the two models may be equivalent close to the quantum critical point and that studies in both these models are relevant to the physics of graphene \cite{herbut:146401,herbut:085116}. The critical exponents in a lattice GN model with a $U_f(1) \times Z_2$ symmetry with staggered fermions have been computed and it was found that $\nu=1.00(4)$ and $\eta = 0.754(8)$ \cite{Karkkainen:1993ef}. These critical exponents have also been obtained from other techniques \cite{PhysRevLett.86.958,PhysRevB.66.205111} and they clearly appear to be different from what we find above. However, since the lattice symmetries are different between the two models, the critical behavior may fall under different universality classes.  The critical behavior with a continuous $U_f(1) \times U_\chi(1)$ GN model was studied in \cite{PhysRevD.53.4616}. It was found that $\nu = 1.02(8)$ and $\beta = 0.89(10)$ which also seem to be different from our results but only at the 2-$\sigma$ level. More work is necessary to clarify the issue of whether the GN model and the Thirring model have the same critical exponents.

\section{Discussion and Conclusions}

In this work we have introduced a new approach to the fermion sign problem which we call the {\em fermion bag} approach. The essential idea behind the new approach is to recognize that fermionic degrees of freedom are usually contained inside dynamically determined space-time regions (bags). Outside these regions they are hidden since they become paired into bosonic objects. Then it is likely that the sign problem is localized to the regions inside these bags and may be solved using determinantal tricks. Such a scenario is clearly natural in systems where there is a coupling $U$ such that when $U$ is infinite all the fermions are paired up and there is no sign problem. Then as $U$ becomes finite, the the fermions are liberated out but are confined to bags. In such systems, if the sign problem is contained within the bags and can be solved, then there will be regions in parameter space where the bags do not grow with the volume. In these regions of parameter space it should be possible to construct Monte Carlo algorithms which are far more efficient than conventional algorithms. In this work we showed an explicit example of a lattice field theory, namely the massless lattice Thirring model, where all these features are realized. In particular we developed a Monte Carlo algorithm and showed that it is efficient for large couplings as expected. For smaller couplings the efficiency of the algorithm was lost, but still the fermion-bag approach continued to provide an alternative approach to the problem. In particular we could determine the critical exponents near the quantum phase transition present in the model with reasonable effort.

While the fermion bag approach loses its main advantage when the bags begin to merge and the fermions are delocalized in the entire space-time volume, it still provides new insights into the sign problem itself. For example we explained in this work, how a natural solution to the {\em silver blaze problem} emerges within this approach. Further, we also showed that new solutions to sign problems emerge. Although the discussion in this work focused on a single model, the idea behind the fermion bag approach is more general and must be applicable to other lattice field theories when formulated cleverly. As an example below we sketch how one can formulate a model to study the physics of the BCS-BEC cross over in the fermion bag approach. 

\begin{figure}[t]
\begin{center}
{
\includegraphics[width=0.7\textwidth]{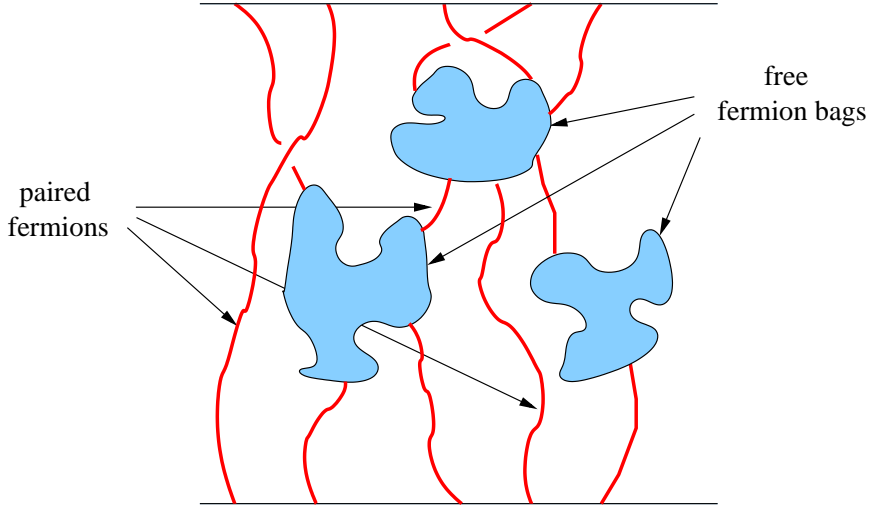}
}
\end{center}
\caption{\label{fig9} A sketch of a fermion bag configuration in a model described in the text that contains the physics of BEC-BCS cross over. The solid lines represent spin-up and spin-down fermions strongly paired into a spin-zero boson. The blobs represent regions where fermions can be liberated and become essentially free.}
\end{figure}

Let $(i,t)$ be the coordinates of a cubic space-time lattice where $i \equiv (i_x,i_y), i_x,i_y = 0,1,2,...,(L-1)$ is the two dimensional spatial coordinate and $t=0,1,2.., (L_t-1)$ is the temporal coordinate. Let $e_i$ represent the four neighboring spatial sites of the site $i$. Let the spin be represented with $s=\uparrow,\downarrow$. The fermion fields are represented by four independent Grassmann variables $\psi_{s,i,t}$ and $\psib_{s,i,t}$ per site which satisfy anti-periodic boundary conditions in time but periodic boundary conditions in space. Using this notation, consider the lattice field theory model described by the action $S = S_0 + U S_1$, where
\begin{subequations}
\begin{eqnarray}
S_0 &=& - 
\sum_{i,t,s} \Bigg\{
\Big(\psib_{s,i,{t+1}}\mathrm{e}^{\mu\varepsilon} - \psib_{s,i,t}\Big) \psi_{s,i,t}\ 
+ \ \varepsilon \ t \ \mathrm{e}^{\mu\varepsilon} \ \sum_{e_i} \psib_{s,e_i,t+1}\psi_{s,i,t} \Bigg\},
\\
S_1 &=& -
\sum_{i,t} \Big\{\mathrm{e}^{2\mu\varepsilon}
\psib_{\downarrow,i,t+1}\psib_{\uparrow,i,t+1}\psi_{\uparrow,i,t}\psi_{\downarrow,i,t}
\ + \ 
\psib_{\downarrow,i,t}\psib_{\uparrow,i,t}\psi_{\uparrow,i,t}\psi_{\downarrow,i,t}
\nonumber \\
&& \hskip2in
+ \ \varepsilon \ \mathrm{e}^{2\mu\varepsilon}\   \sum_{e_i} 
\psib_{\downarrow,e_i,t+1}\psib_{\uparrow,e_i,t+1}\psi_{\uparrow,i,t}\psi_{\downarrow,i,t}
\Big\}.
\end{eqnarray}
\end{subequations}
Note that the action is invariant under $SU(2)$ spin symmetry and $U(1)$ fermion number symmetry as required.

The partition function is given by
\begin{equation}
Z = \int \prod_{x,t,s} [d\psib_{x,t,s}\ d\psi_{x,t,s}] \ \exp(-S).
\end{equation}
 When $U=\infty$ the model describes the physics of paired fermions (hard-core bosons) hopping on the lattice, similar to the quantum XY model. On the other hand when $U=0$, the fermions are free. Hence, as we tune $U$, the model should describe the physics of BEC-BCS crossover. For intermediate values of $U$ we expect regions where fermions are paired and regions where they are free. A sketch of such a configuration is shown in Fig.~\ref{fig9}. The regions where the fermions are free are shown in as a bag in the figure. Further, it is easy to argue that the Boltzmann weight is always positive due to two flavor nature of the problem.

Based on the above reasoning we believe we have uncovered a somewhat unconventional approach to fermionic field theories which may prove to be a powerful alternative, especially when the coupling strengths are large and where perturbation theory is expected to fail.

\section*{Acknowledgments}

I thank  Sourendu Gupta for discussions about the HMC algorithm and providing hospitality at TIFR where part of this work was done, Simon Hands for discussions about the Thirring model, Matt Hastings for discussions into the applicability of the fermion bag approach to non-relativistic fermions and Uwe-Jens Wiese for insight about the general nature of the bag approach. This work was supported in part by the Department of Energy grant DE-FG02-05ER41368 and the National Science Foundation grant DMR-0506953.

\bibliography{ref}

\clearpage

\appendix

\section{Algorithm versus Exact Results}

In this section we present some exact results on a $2^3$ lattice and compare them with the results from the algorithm in order to test the algorithm. Table \ref{tab:zconf} gives the various possible configurations (their degeneracy factors (Deg), the corresponding bag determinants (Bdet) and the Boltzmann weights (BWt)).
\begin{table}[ht]
\begin{tabular}{|c|c|c|c||c|c|c|c|}
\hline\hline
Config. & Deg. & Bdet. & BWt & Config. & Deg. & Bdet. & BWt \\
\hline 
&&&&&&& \\
\begin{minipage}[c]{0.2\textwidth}
\includegraphics[width=0.3\textwidth]{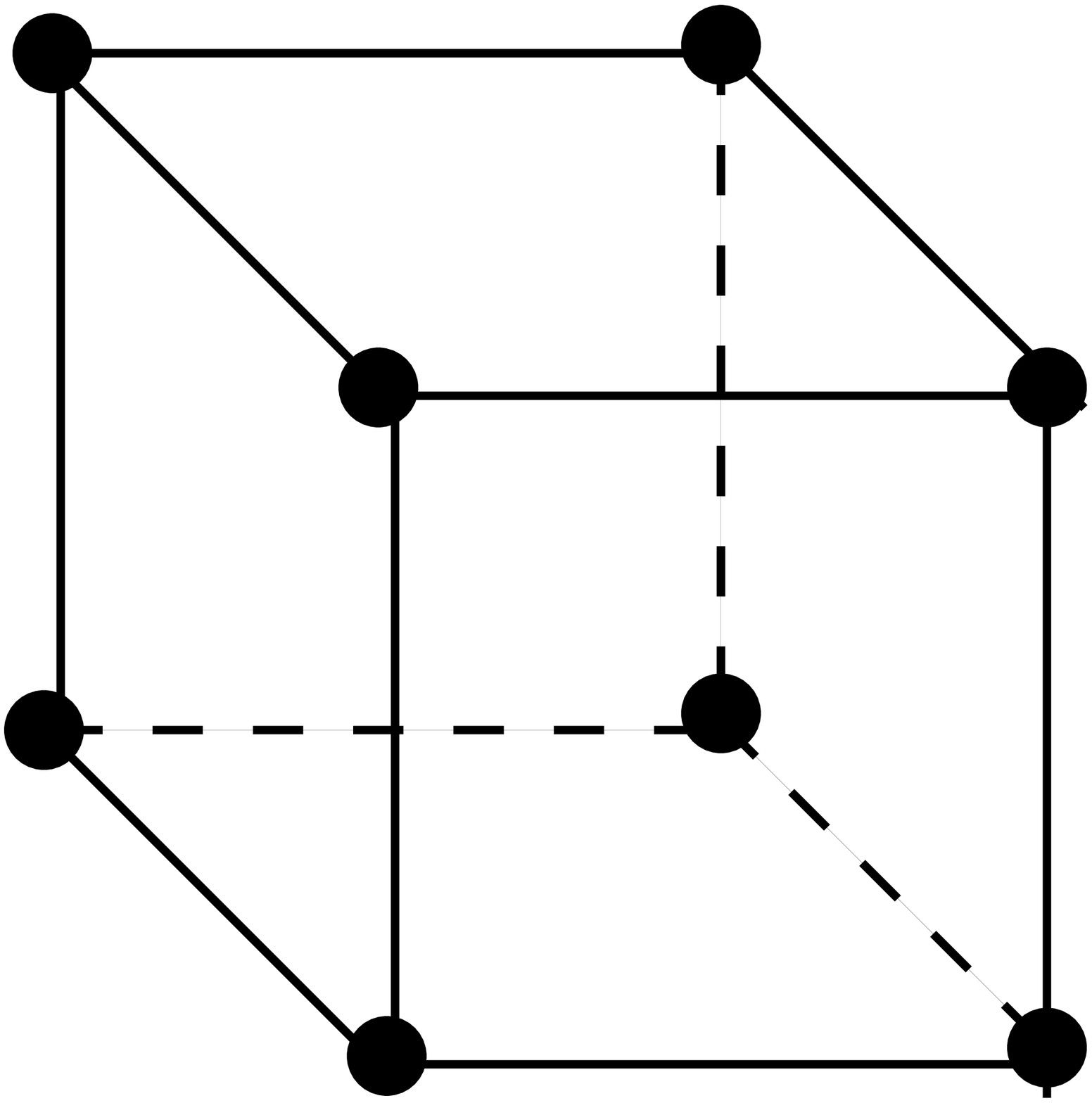}
\end{minipage}
& 1 & 81 & 81 & 
\begin{minipage}[c]{0.2\textwidth}
\includegraphics[width=0.3\textwidth]{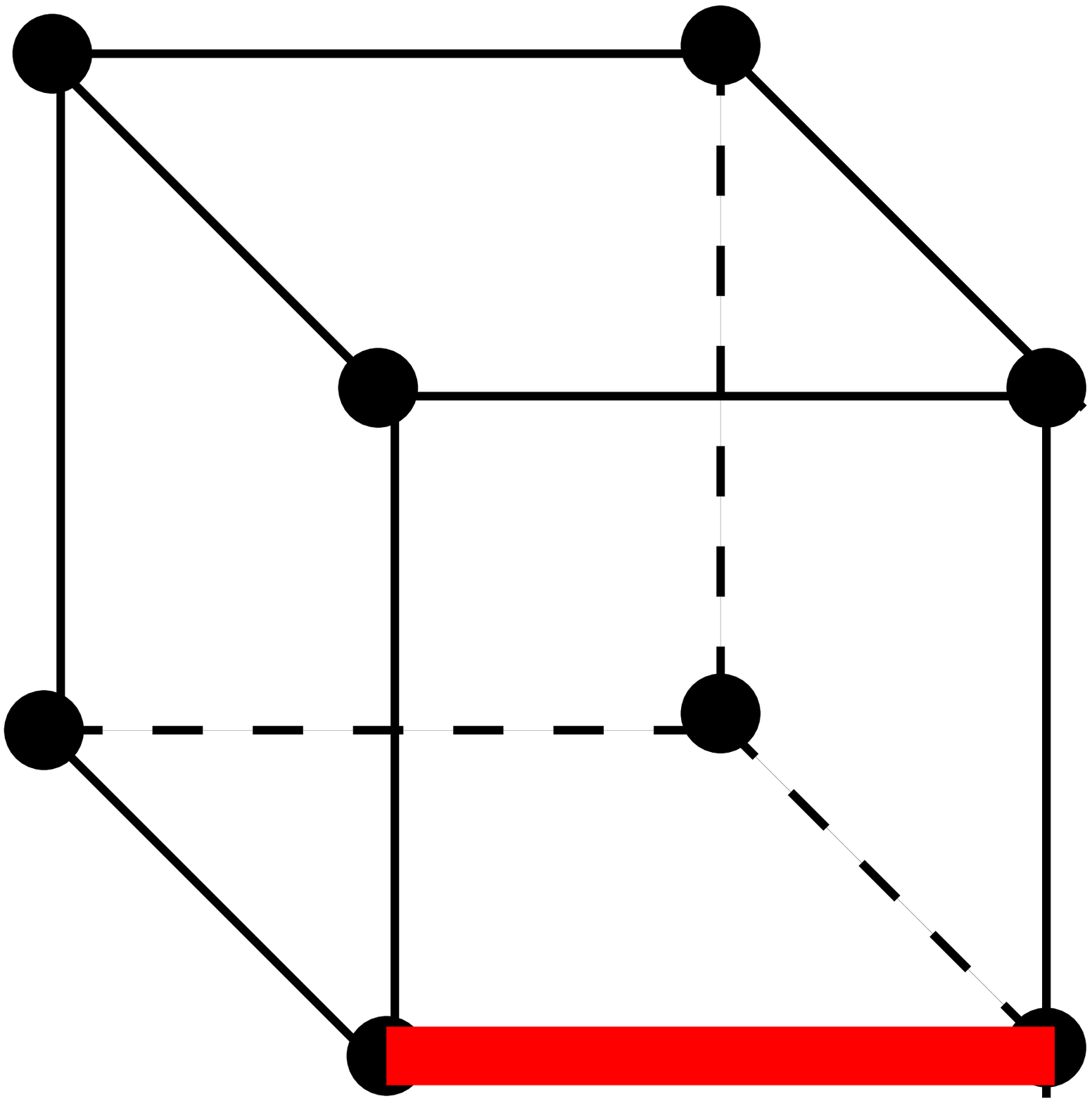}
\end{minipage}
& 24 & 9 & $216 U$ \\
&&&&&&& \\
\begin{minipage}[c]{0.2\textwidth}
\includegraphics[width=0.3\textwidth]{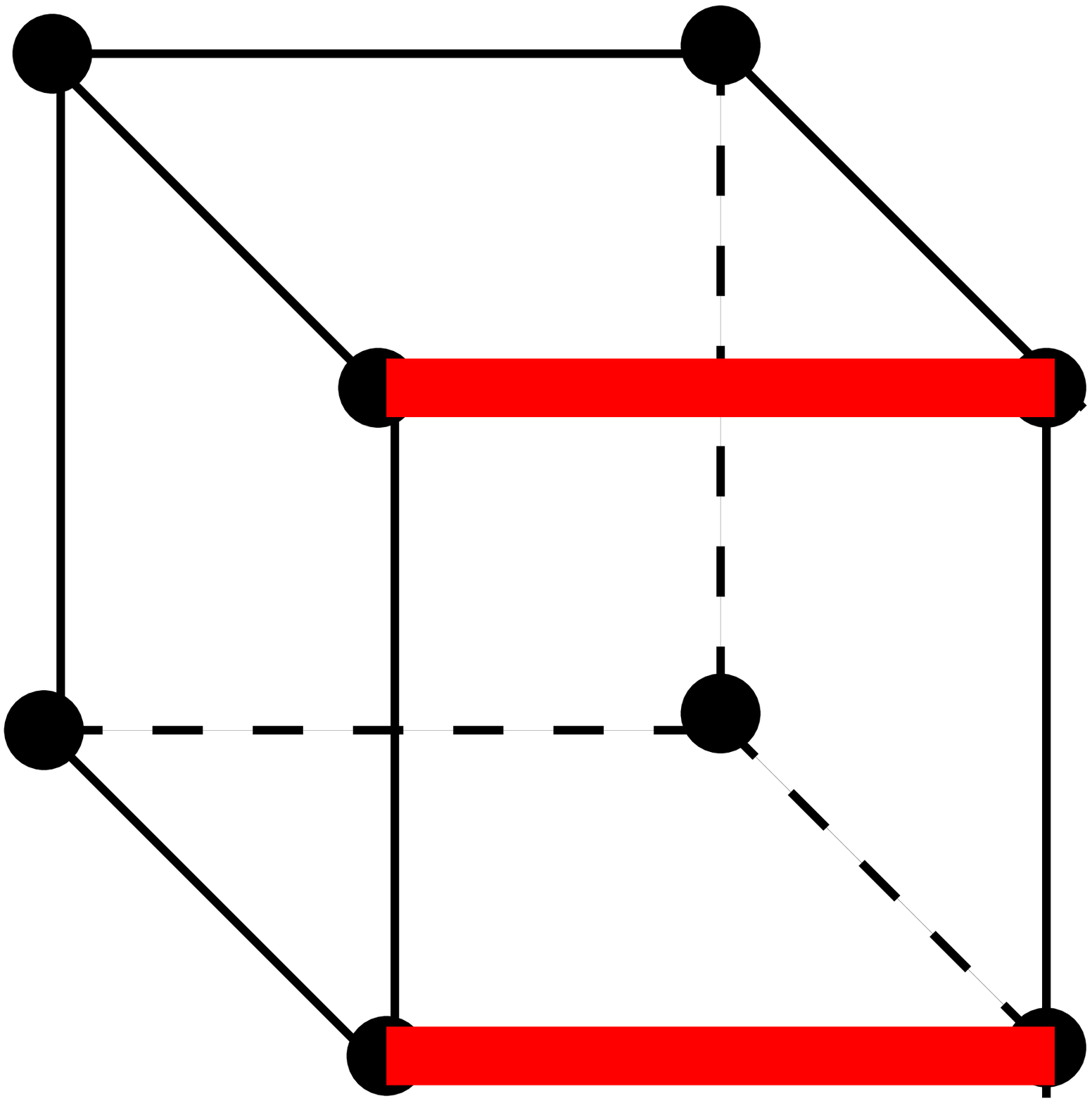}
\end{minipage}
& 48 & 4 & $192 U^2$ & 
\begin{minipage}[c]{0.2\textwidth}
\includegraphics[width=0.3\textwidth]{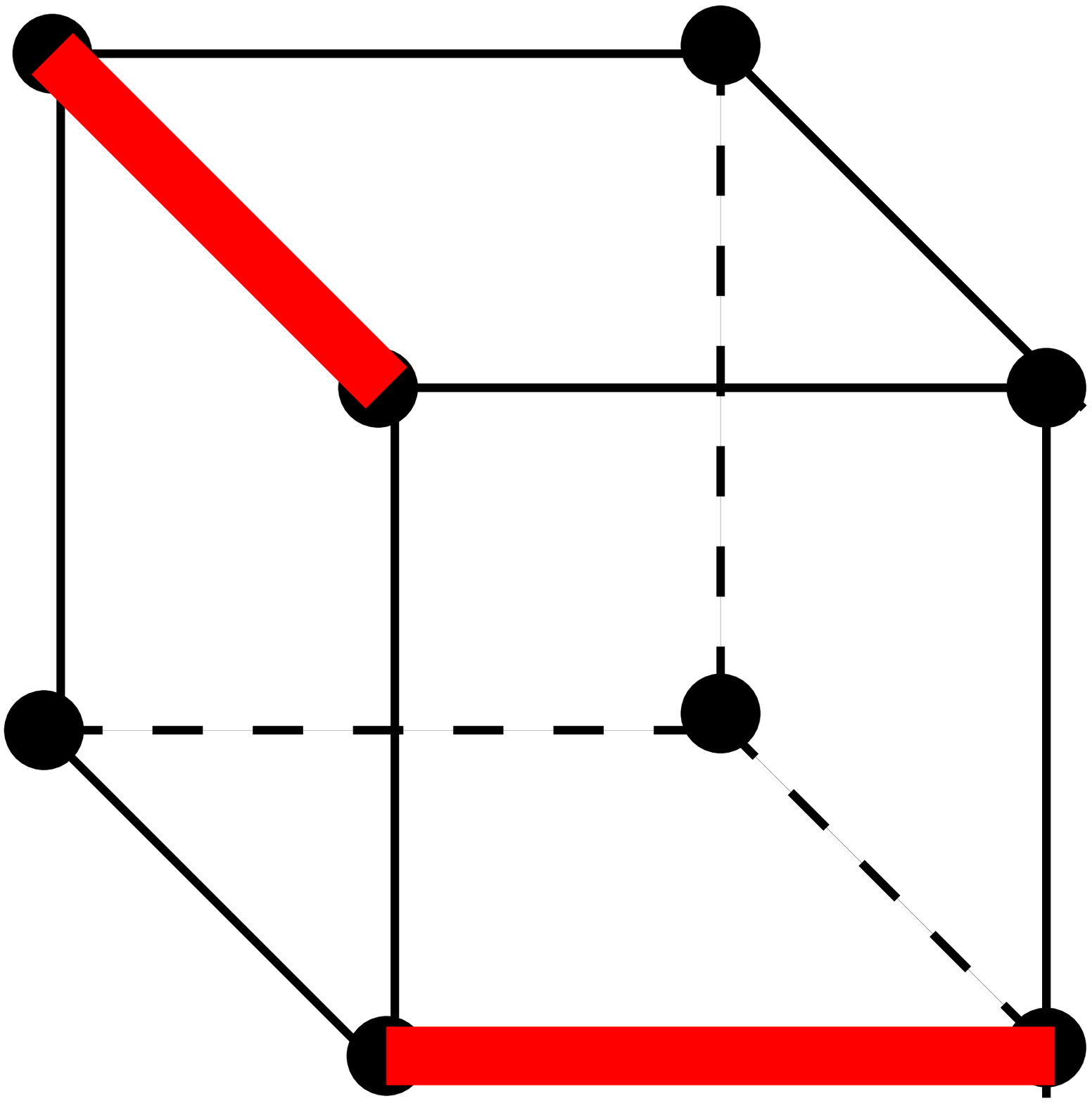}
\end{minipage}
& 96 & 1 & $96 U^2$ \\
&&&&&&& \\
\begin{minipage}[c]{0.2\textwidth}
\includegraphics[width=0.3\textwidth]{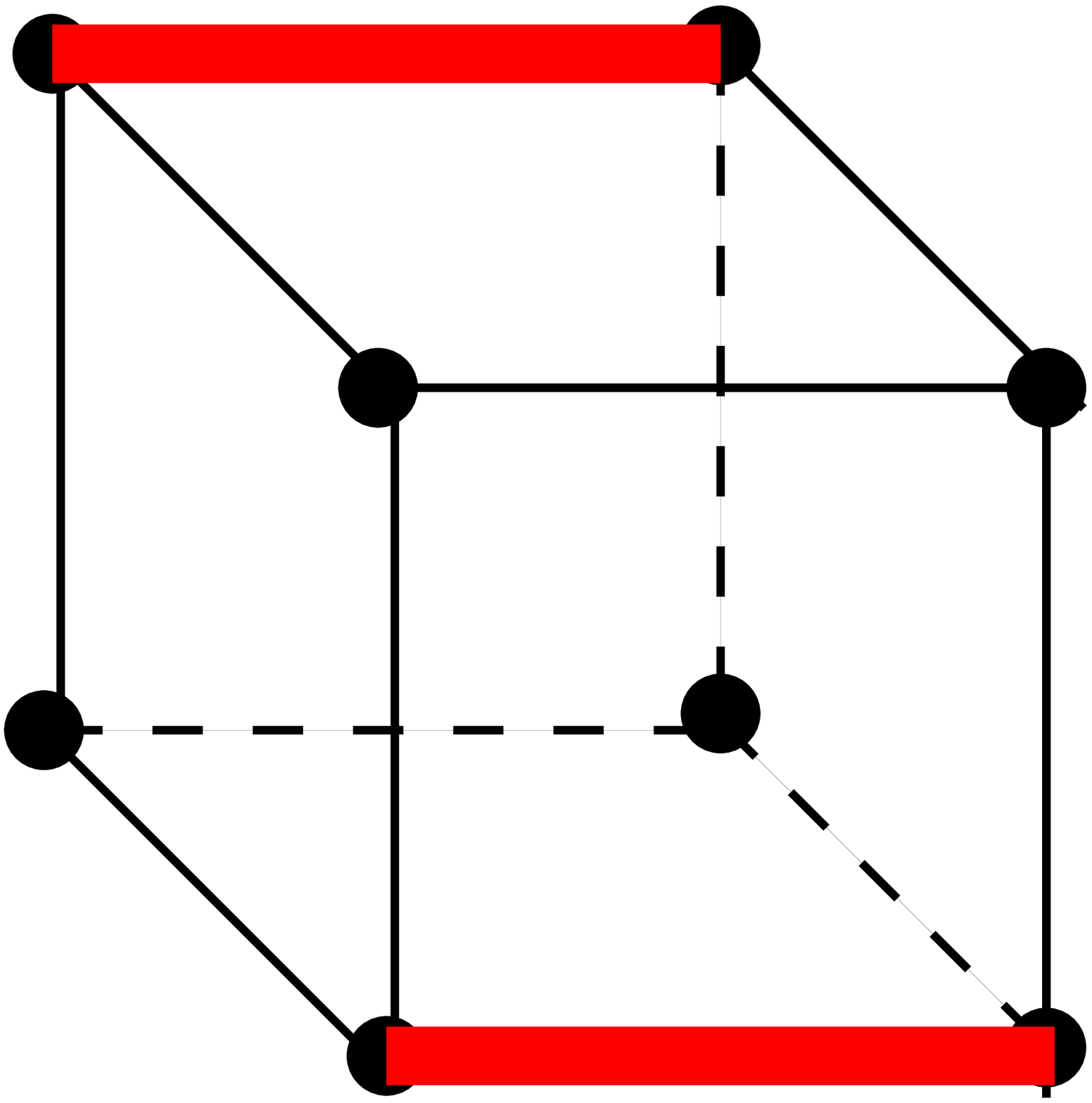}
\end{minipage}
& 24 & 1 & $24 U^2$ &
\begin{minipage}[c]{0.2\textwidth}
\includegraphics[width=0.3\textwidth]{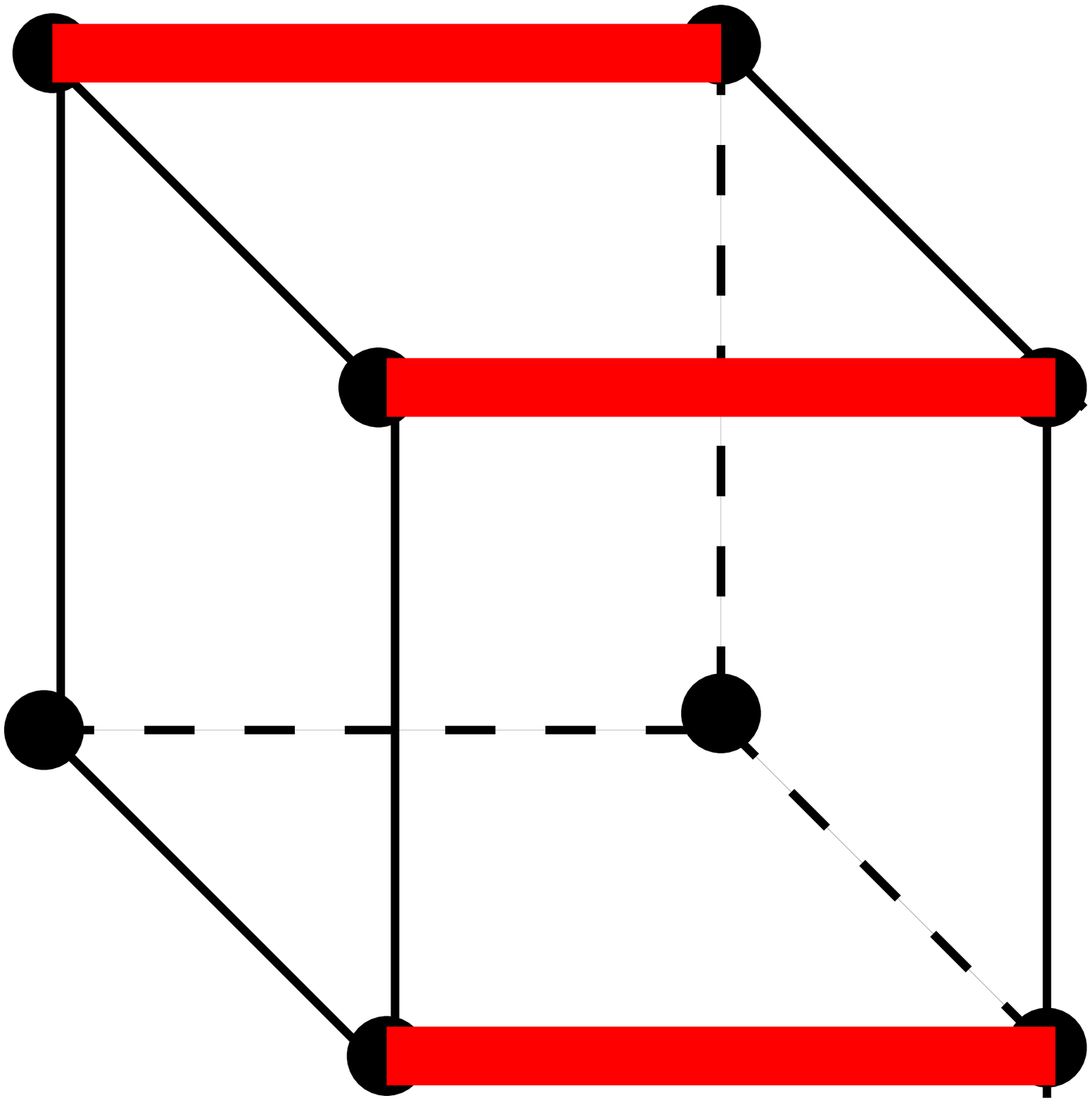}
\end{minipage}
& 96 & 1 & $96 U^3$ \\
&&&&&&& \\
\begin{minipage}[c]{0.2\textwidth}
\includegraphics[width=0.3\textwidth]{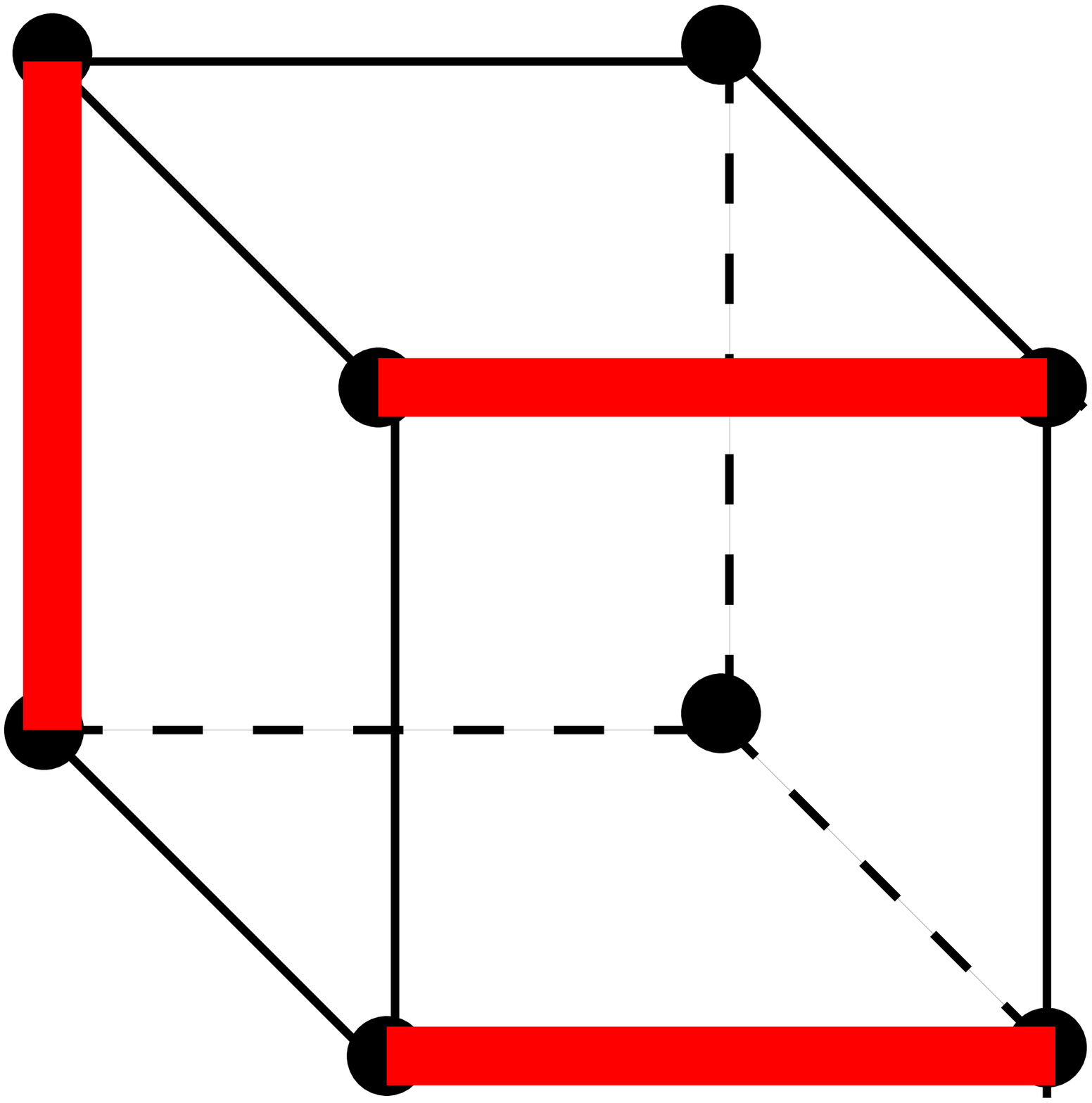}
\end{minipage}
& 192 & 1 & $192 U^3$ &
\begin{minipage}[c]{0.2\textwidth}
\includegraphics[width=0.3\textwidth]{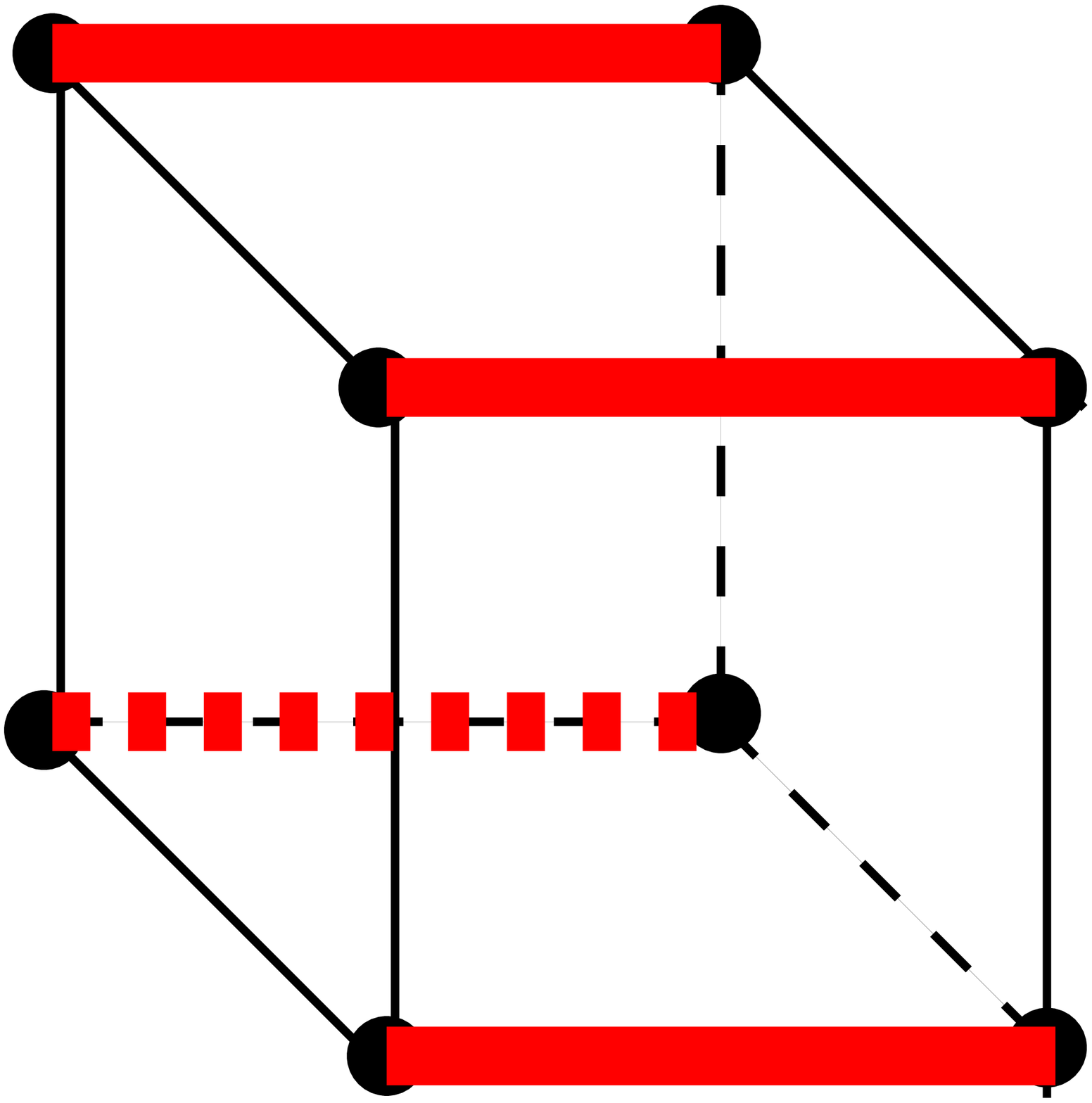}
\end{minipage}
& 48 & 1 & $48 U^4$ \\
&&&&&&& \\
\begin{minipage}[c]{0.2\textwidth}
\includegraphics[width=0.3\textwidth]{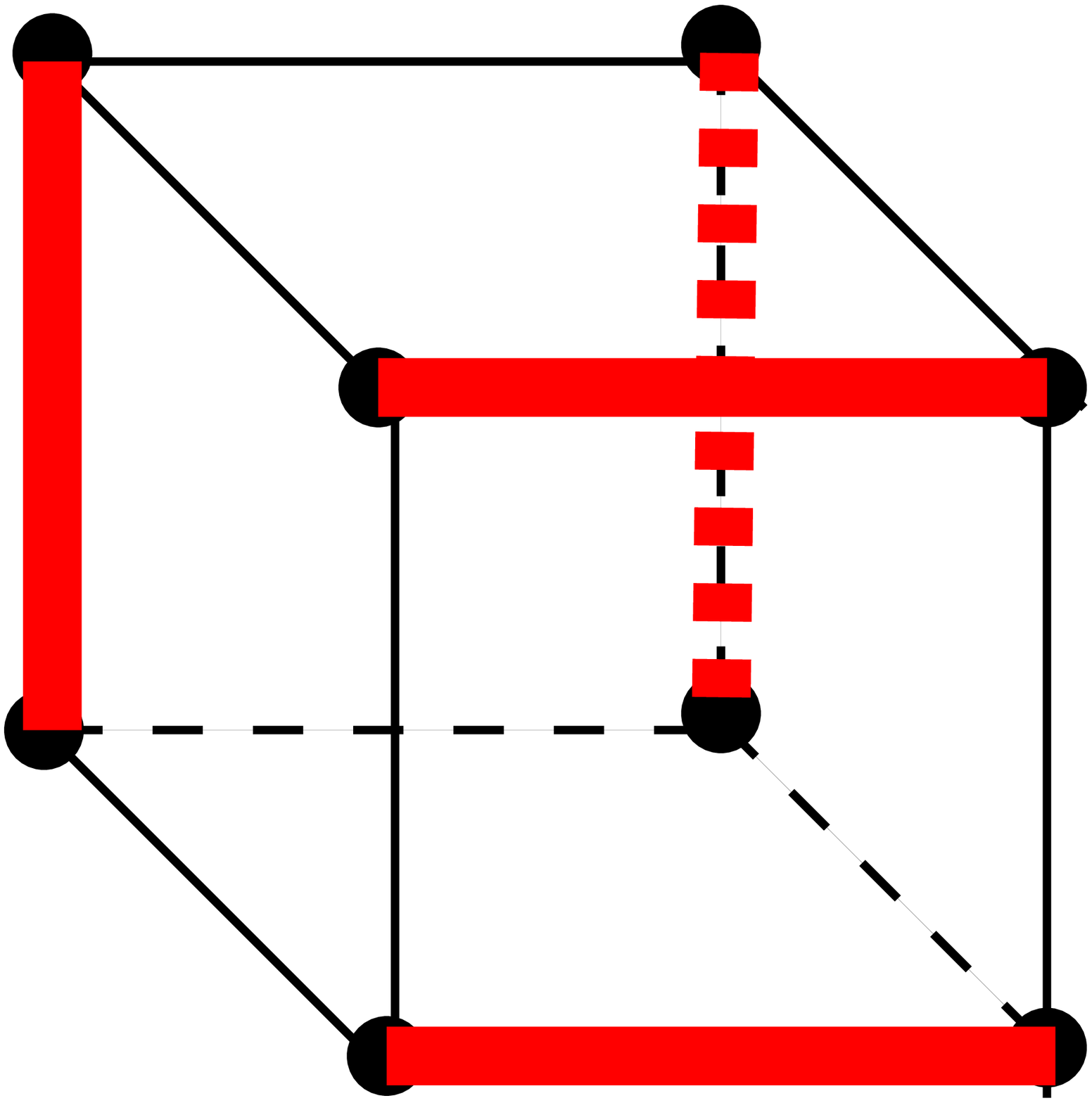}
\end{minipage}
& 96 & 1 & $96 U^4$ & & & & \\
&&&&&&& \\ 
\hline
\hline
\end{tabular}
\caption{\label{tab:zconf} Contributions to the partition function for $2^3$ lattice.}
\end{table}
We find that the partition function is given by
\begin{equation}
Z = 81 + 216 U + 312 U^2 + 288 U^3 + 144 U^4
\end{equation}
The average number of bonds is given by
\begin{equation}
\langle N_B \rangle = \frac{1}{4 Z} \ ( 216 U + 624 U^2 + 864 U^3 + 576 U^4)
\end{equation}
where we have normalized it so that for large $U$ it approaches one. The typical bag size is given by
\begin{equation}
\langle S_\tau \rangle = \frac{1}{Z} \ (648 + 972 U + 600 U^2 + 144 U^3)
\end{equation}
The condensate susceptibility is obtained from configurations with two monomers. These along with their degeneracy factors, bag determinants and Boltzmann weights are given in table \ref{tab:psiconf}.
\begin{table}[ht]
\begin{tabular}{|c|c|c|c||c|c|c|c|}
\hline\hline
Config. & Deg. & Bdet. & BWt & Config. & Deg. & Bdet. & BWt \\
\hline 
&&&&&&& \\
\begin{minipage}[c]{0.2\textwidth}
\includegraphics[width=0.3\textwidth]{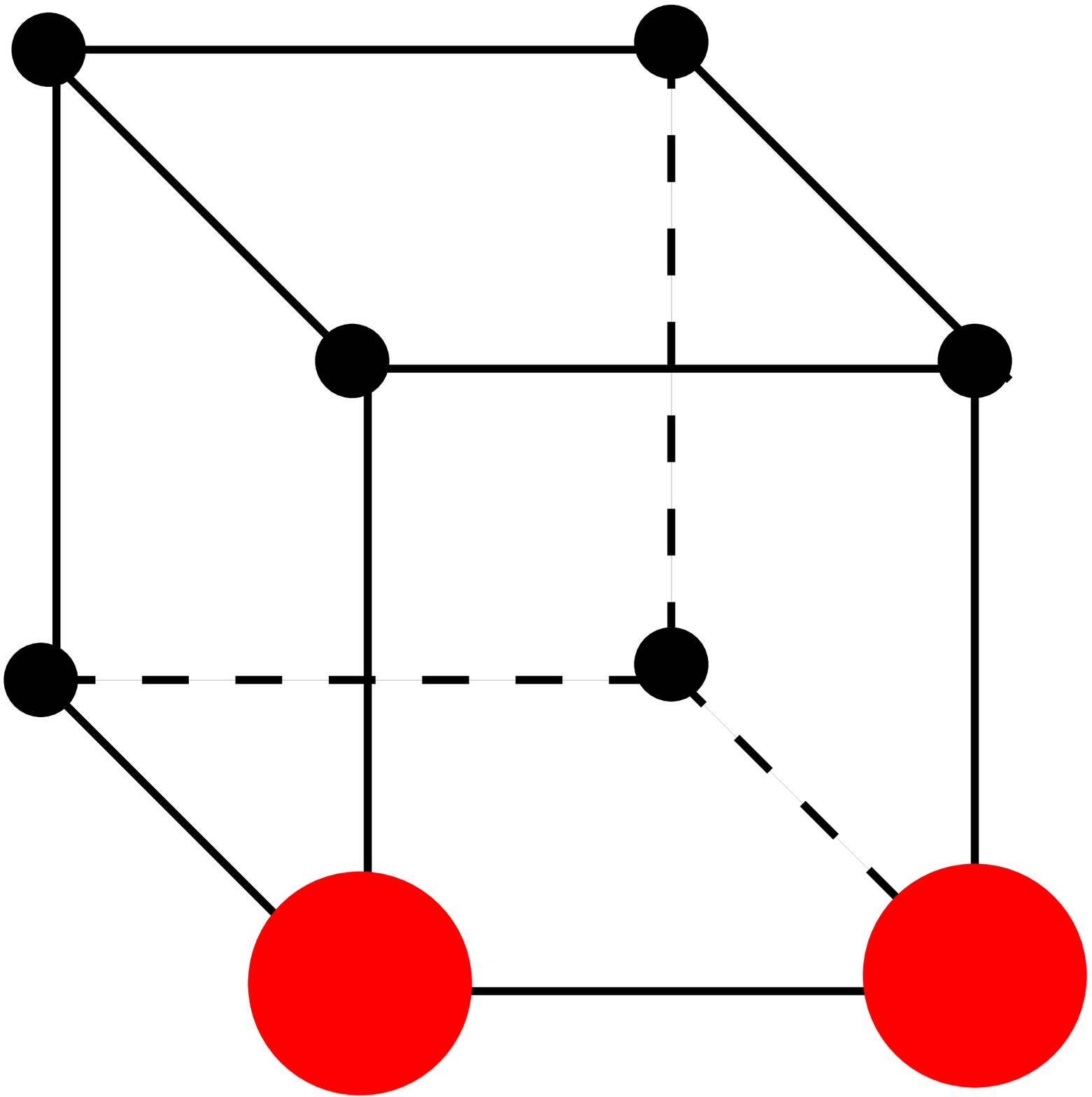}
\end{minipage}
& 3 & 9 & 27 
&
\begin{minipage}[c]{0.2\textwidth}
\includegraphics[width=0.3\textwidth]{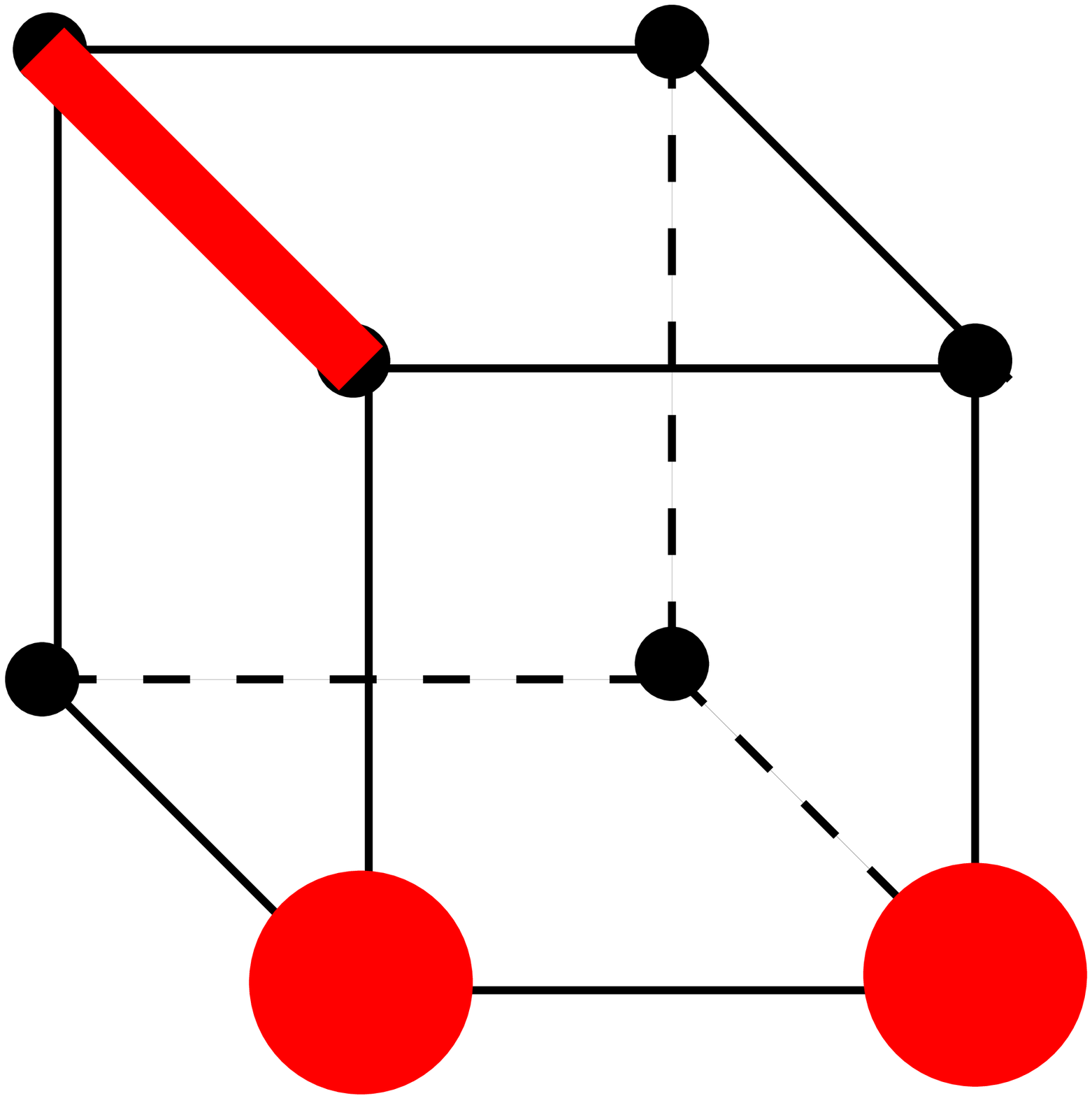}
\end{minipage}
& 24 & 1 & $24 U$ \\
&&&&&&& \\
\begin{minipage}[c]{0.2\textwidth}
\includegraphics[width=0.3\textwidth]{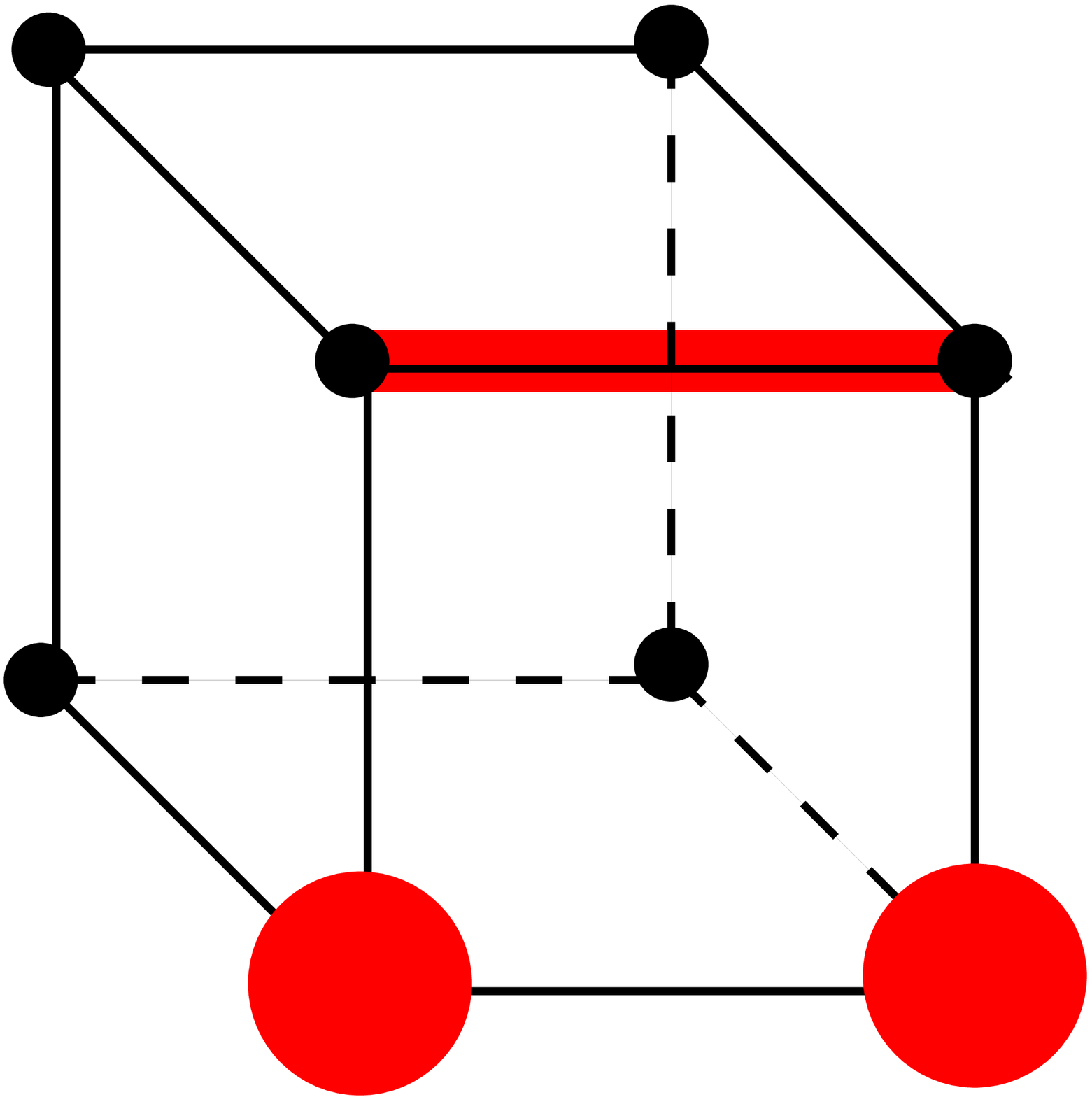}
\end{minipage}
& 12 & 4 & $48 U$ &
\begin{minipage}[c]{0.2\textwidth}
\includegraphics[width=0.3\textwidth]{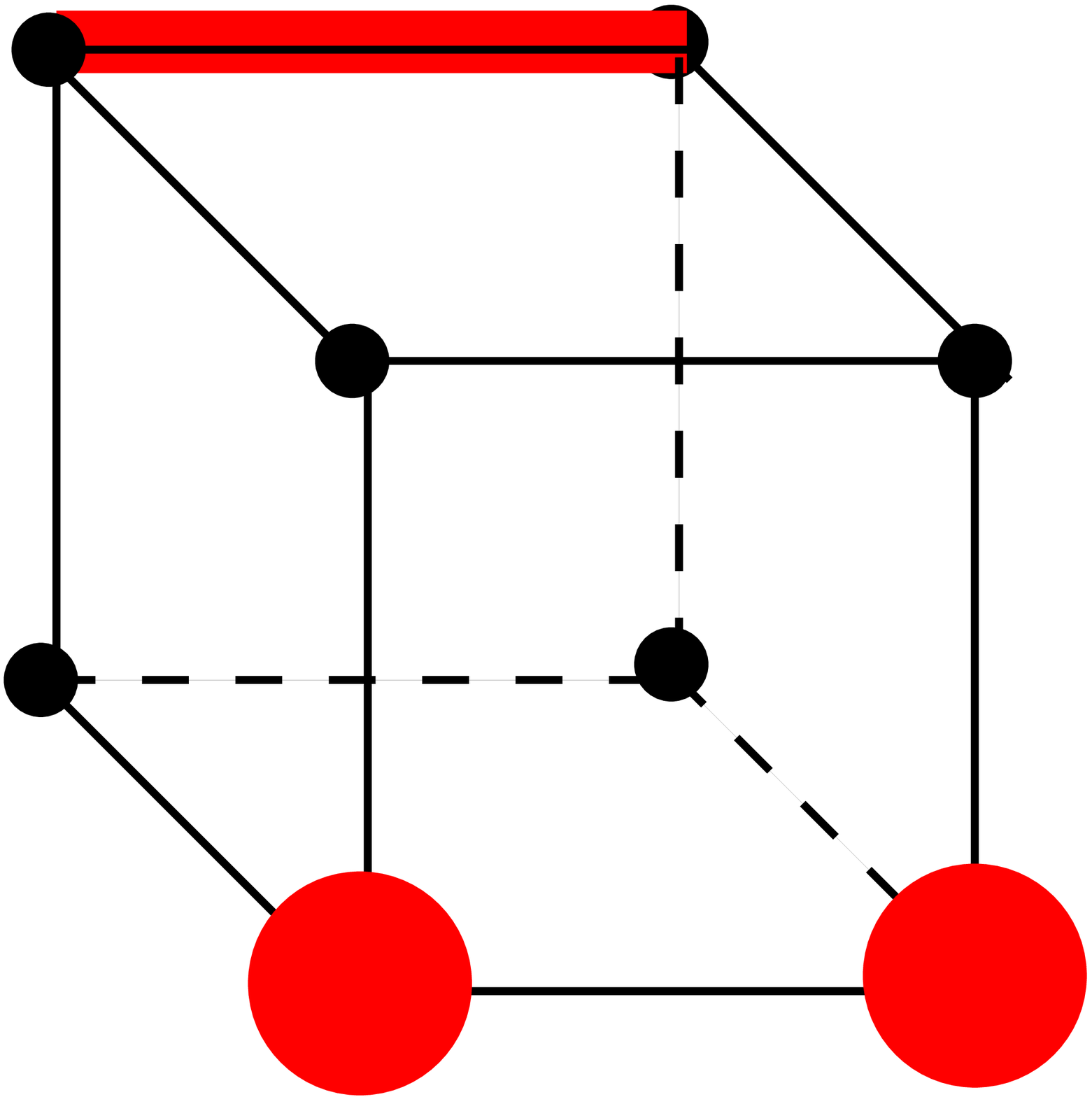}
\end{minipage}
& 6 & 1 & $6 U$ \\
&&&&&&& \\
\begin{minipage}[c]{0.2\textwidth}
\includegraphics[width=0.3\textwidth]{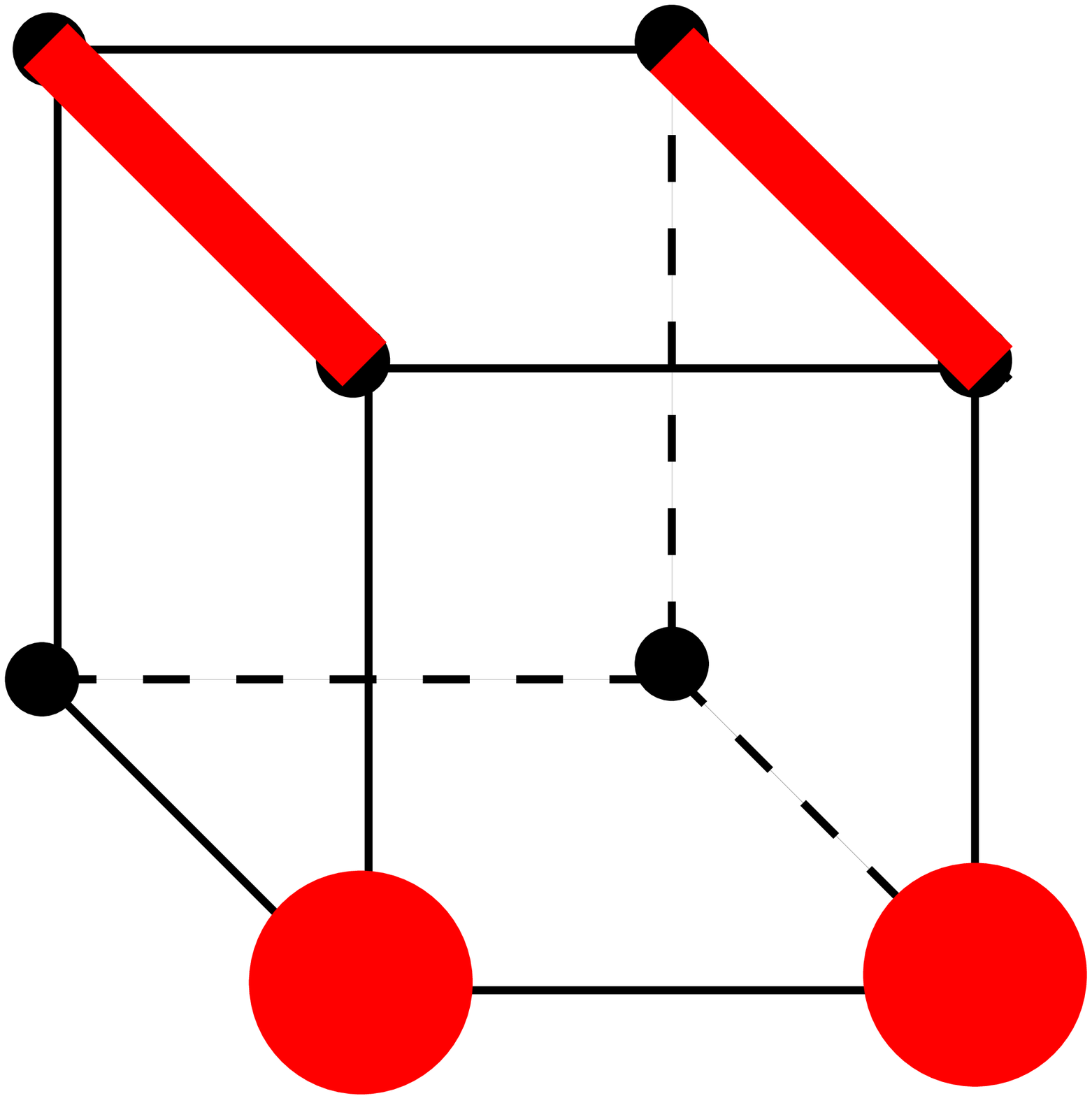}
\end{minipage}
& 108 & 1 & $108 U^2$ &
\begin{minipage}[c]{0.2\textwidth}
\includegraphics[width=0.3\textwidth]{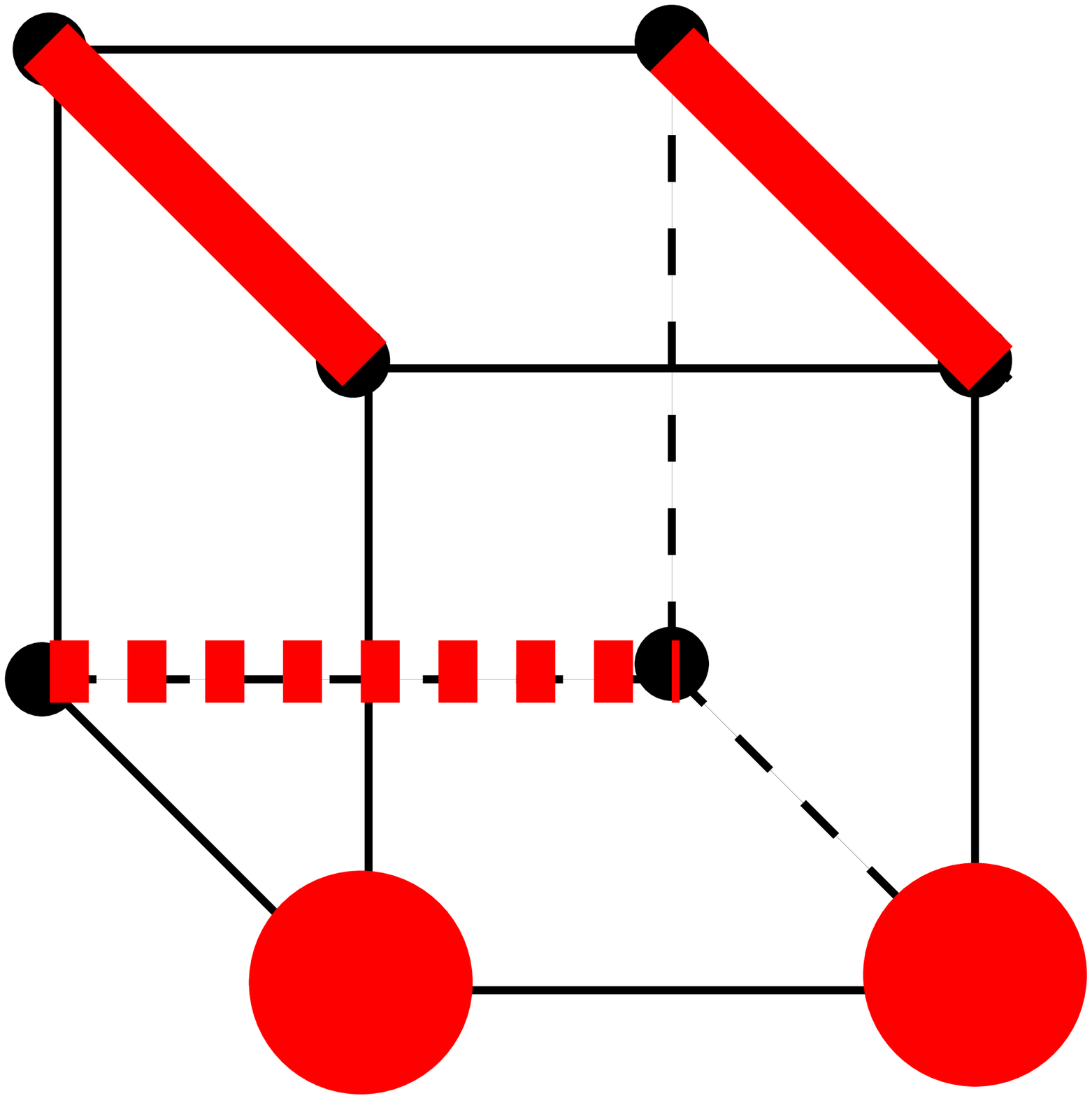}
\end{minipage}
& 72 & 1 & $72 U^3$ \\
&&&&&&& \\
\begin{minipage}[c]{0.2\textwidth}
\includegraphics[width=0.3\textwidth]{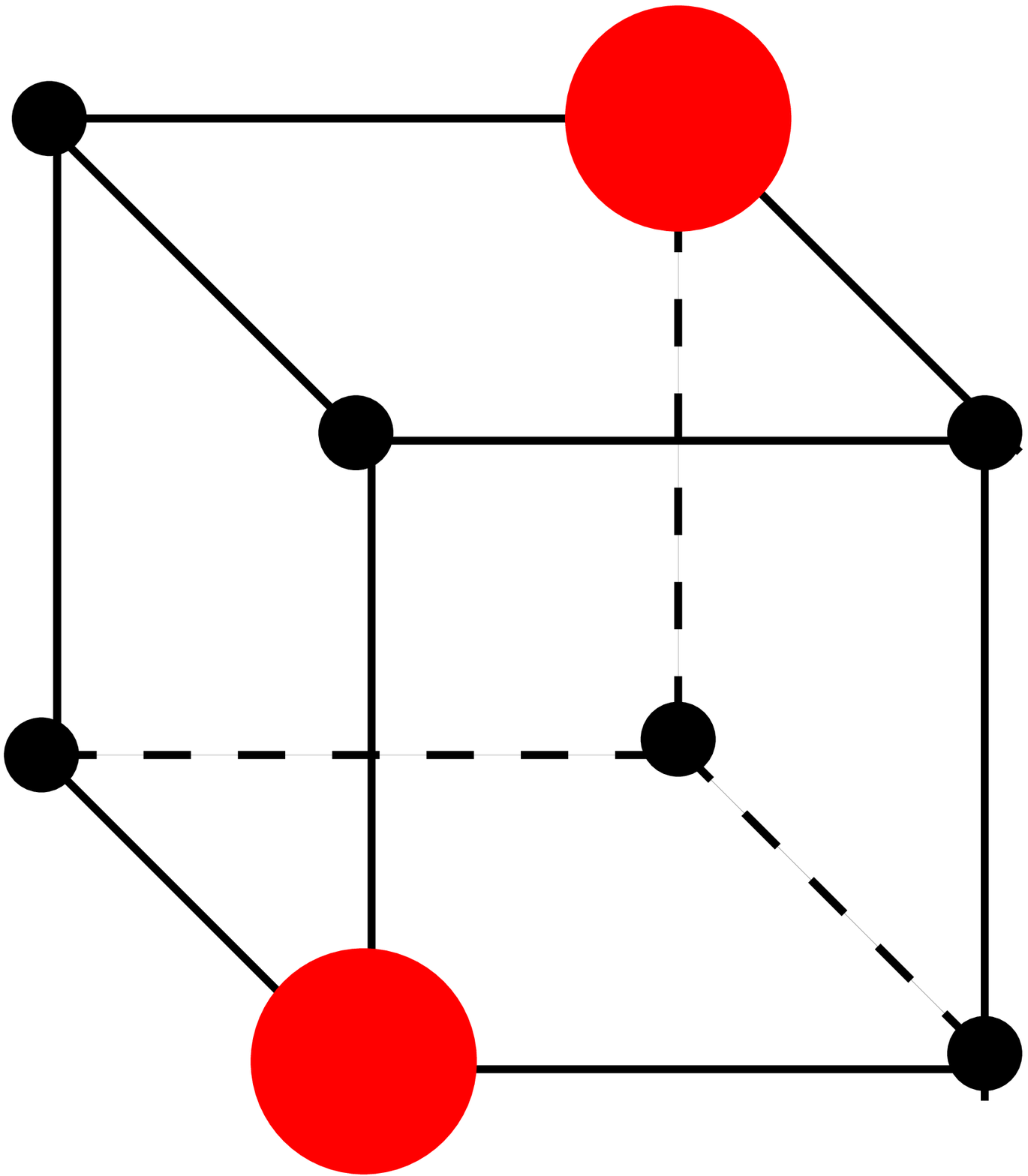}
\end{minipage}
& 1 & 0 & 0 &
\begin{minipage}[c]{0.2\textwidth}
\includegraphics[width=0.3\textwidth]{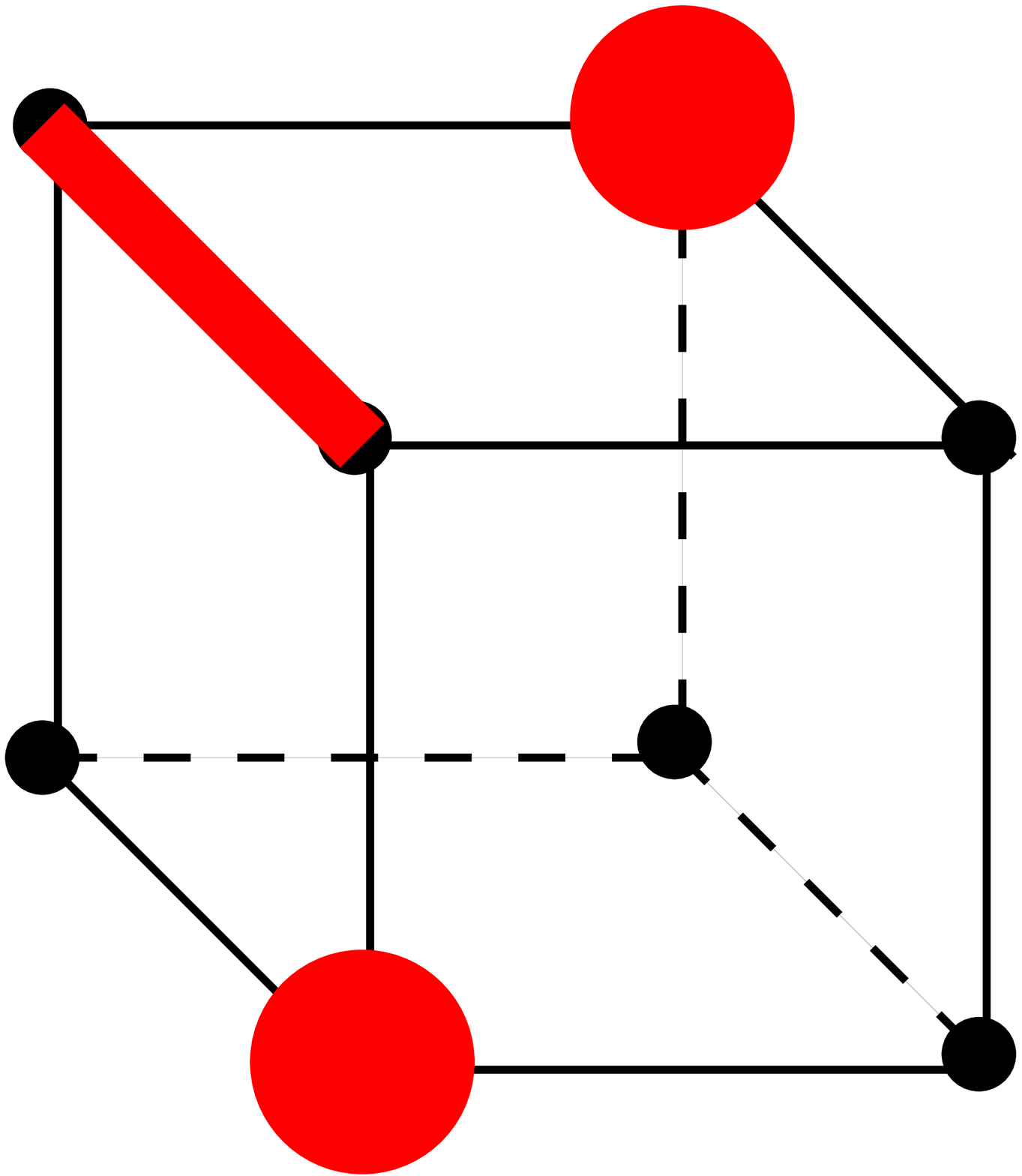}
\end{minipage}
& 12 & 1 & $12 U$ \\
&&&&&&& \\
\begin{minipage}[c]{0.2\textwidth}
\includegraphics[width=0.3\textwidth]{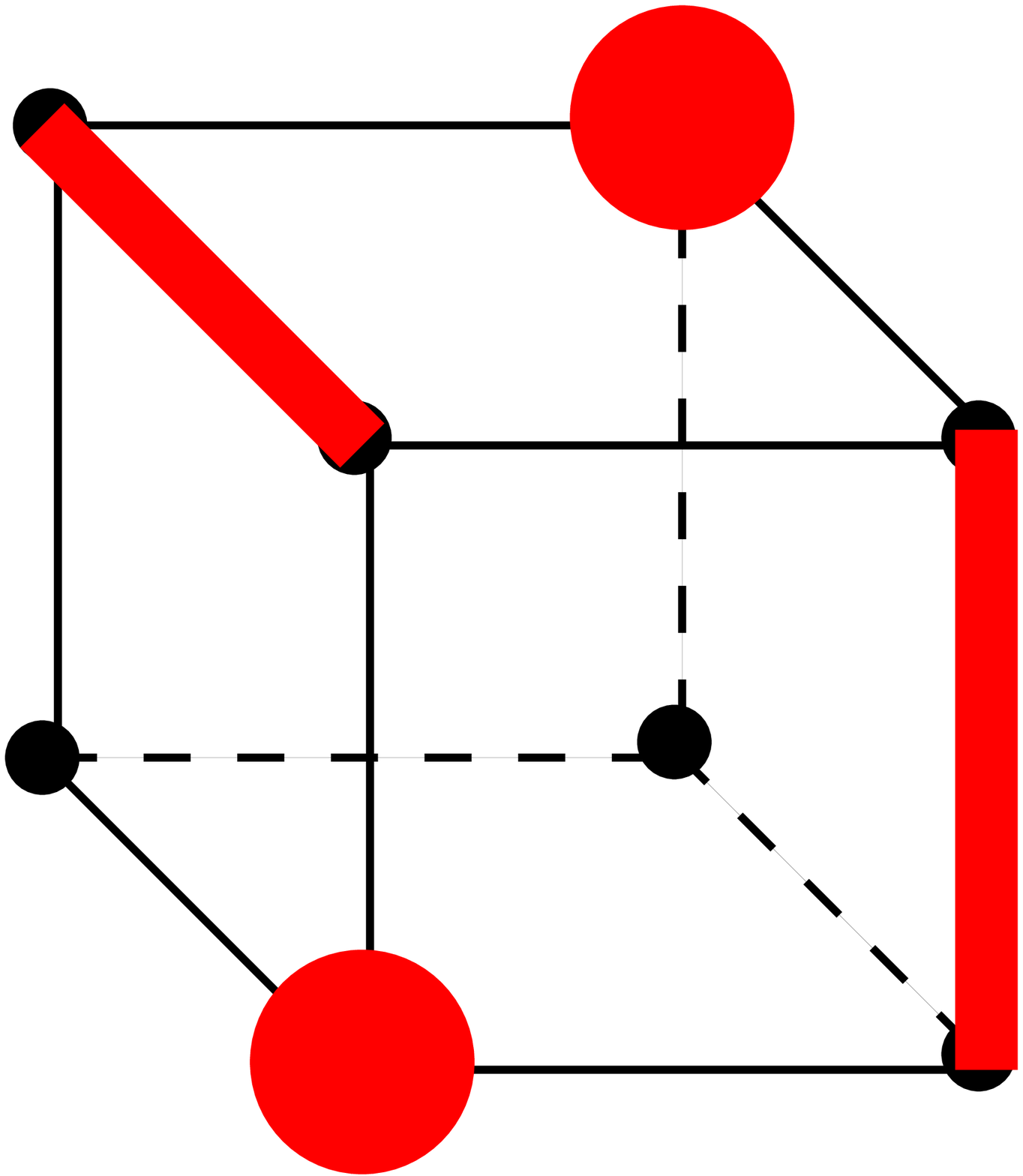}
\end{minipage}
& 24 & 1 & $24 U^2$ &
\begin{minipage}[c]{0.2\textwidth}
\includegraphics[width=0.3\textwidth]{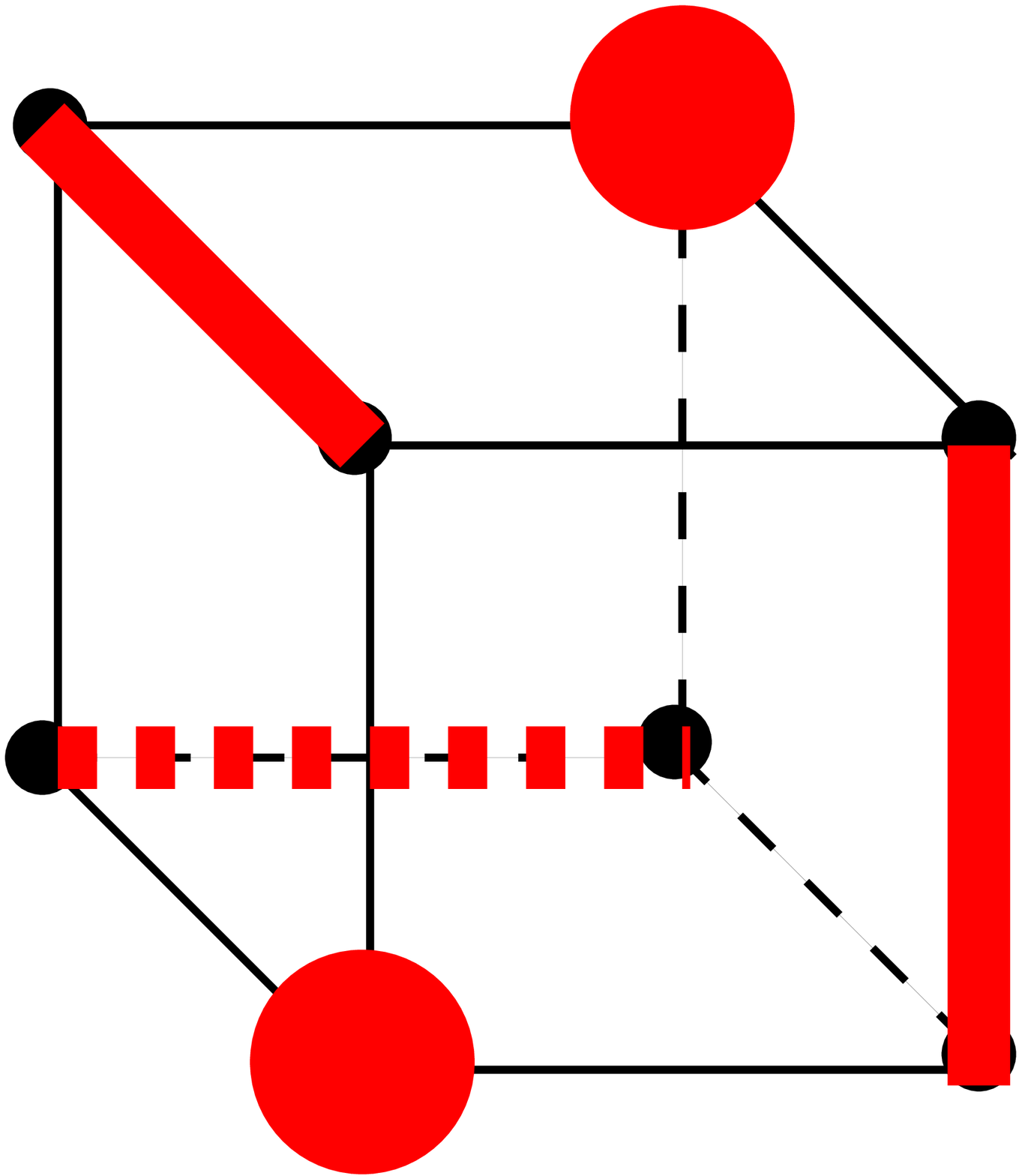}
\end{minipage}
& 16 & 1 & $16 U^3$ \\
&&&&&&& \\
\hline
\hline
\end{tabular}
\caption{\label{tab:psiconf} Contributions to the condensate susceptibility.}
\end{table}
Based on this table we find that
\begin{equation}
\chi = \frac{U}{Z} (27 + 90 U + 132 U^2 +  88 U^3)
\end{equation}
The above exact expressions have been tested against our Monte Carlo method. The results for a few values of $U$ are shown in table ~\ref{tab:compare}.
\begin{table}[h]
\begin{center}
\begin{tabular}{|c||c|c||c|c||c|c|}
\hline
U & \multicolumn{2}{|c||}{$N_B$} & \multicolumn{2}{|c||}{$S_\tau$} & \multicolumn{2}{|c|}{$\chi$} \\
\hline
& Exact & Monte Carlo & Exact & Monte Carlo & Exact & Monte Carlo \\
\hline
0.1 & 0.06781... & 0.0678(1) & 7.0866... & 7.087(1) & 0.35283... & 0.3514(9) \\
0.5 & 0.32692... & 0.3270(3) & 4.1730... & 4.173(2) & 0.37179... & 0.3721(5) \\
1.0 & 0.54755... & 0.5477(3) & 2.2708... & 2.269(2) & 0.32372... & 0.3238(3) \\
3.0 & 0.82961... & 0.8298(4) & 0.5593... & 0.559(1) & 0.16803... & 0.1682(1) \\
\hline
\end{tabular}  
\end{center}
\caption{\label{tab:compare} Comparison between exact results and results from the Monte Carlo algorithm for the three observables $N_B$, $S_\tau$ and $\chi$.}
\end{table}

\end{document}